\pdfoutput=1
\documentclass[preprint,11pt,times]{elsarticle}
\usepackage[margin=1in]{geometry}
\usepackage{amsmath,amssymb,amsthm,mathtools,bm,booktabs,graphicx,enumitem,array}
\usepackage[colorlinks=true,linkcolor=blue,citecolor=blue,urlcolor=blue]{hyperref}
\usepackage[expansion=false]{microtype}
\journal{Journal of Multivariate Analysis}
\newtheorem{theorem}{Theorem}
\newtheorem{proposition}{Proposition}
\newtheorem{lemma}{Lemma}
\newtheorem{assumption}{Assumption}

\newtheorem{remark}{Remark}
\newcommand{\R}{\mathbb R}\newcommand{\E}{\mathbb E}
\newcommand{\Var}{\operatorname{Var}}\newcommand{\Cov}{\operatorname{Cov}}
\newcommand{\indep}{\perp\!\!\!\perp}\newcommand{\I}{\mathbf 1}
\newcommand{\tp}{^{\mathsf T}}\newcommand{\Acal}{\mathcal A}
\newcommand{\Hcal}{\mathcal H}
\newcommand{\Fcal}{\mathcal F}
\newcommand{\Pn}{\mathbb P_n}
\newcommand{\Csp}{\mathcal C^{\rm sp}}
\begin{document}
\begin{frontmatter}
\title{Exhaustive adjustment-set selection and causal effects for distribution-valued outcomes, with inference under latent Gaussian spatio-temporal dependence\tnoteref{tsupp}}
\tnotetext[tsupp]{References to Sections S1--S7, Tables S1--S10 and Figure S1 are to the Supplementary Material, which is available from the author on request.}
\author{Hsin-Hsiung Huang\corref{cor}}
\ead{hsin.huang@ucf.edu}
\cortext[cor]{Corresponding author.}
\address{School of Data, Mathematical, and Statistical Sciences, University of Central Florida, Orlando, FL 32816, USA}
\begin{abstract}
Selecting adjustment covariates for a distribution-valued outcome must preserve dependence that scalar summaries can discard, and surveillance outcomes are also dependent across space and time. Under unconfoundedness and a relative-overlap condition we identify average treatment effects, stochastic and incremental intervention effects, and natural direct and indirect effects for such outcomes, with doubly and triply robust scores. For independent data, adjustment sets are selected exhaustively by distance-covariance sufficient dimension reduction: we give a root-$n$ projector expansion, a basis-invariant criterion tensor, and a simultaneous outer confidence collection calibrated by fractional sampling with a finite-population correction. For dependence, units are functions of a latent Gaussian field, exponential in space and Ornstein--Uhlenbeck in time, plus independent noise. Such data are $\rho$-mixing at an exponential rate, and the covariance between any two square-integrable scores is a power series in the latent correlation with positive semidefinite coefficients. Consequences include a central limit theorem for all the causal estimators, the exact undercoverage of independence-based intervals, a consistent Mehler covariance estimator under finite chaos order, and an asymptotically conservative convex-chord estimator; block fractional sampling extends the collection under an assumed linearization. In simulations, fractional sampling raises weak-signal collection coverage from $.69$--$.78$ to at least $.98$ while retaining nearly all candidate sets, and the dependence-aware estimators substantially reduce undercoverage, although none is exact when few units fall within a correlation range. A descriptive analysis of U.S.\ syphilis surveillance and Medicaid expansion illustrates the estimands and compares dependence-aware with conventional inference.
\end{abstract}
\end{frontmatter}

\section{Introduction}
A distribution-valued response can depend on covariates through location, scale or shape, so an adjustment set adequate for its mean need not remove confounding for the distribution itself. For example, the average time of incidence can be unchanged while its temporal concentration differs between exposure groups. Public-health surveillance adds a second difficulty: jurisdictions are neighbours, policies cluster geographically and incidence is persistent, so the independent-sampling approximations behind standard errors and resampling calibrations can fail.

\citet{luo2024exhaustive} introduced exhaustive identification of sufficient adjustment sets through outcome and treatment central subspaces; we build on that criterion and do not propose exhaustive selection as new. Distance-covariance sufficient dimension reduction (SDR) \citep{sheng2013direction,sheng2016sufficient,huang2024rsdr,huang2025fsdr} supplies natural estimators when scalar inverse-regression summaries are inadequate. Its root-$n$ theory for ordinary responses is due to \citet{sheng2013direction,sheng2016sufficient}. Our extension concerns powered and feature-distance responses, a whitening-aware projector expansion, and simultaneous inference on the adjustment collection. The contributions are as follows.
\begin{enumerate}[label=(\roman*),leftmargin=2em,itemsep=1pt]
\item \emph{A basis-invariant criterion tensor and its calibration.} The criterion is written through products of projectors and residual covariance matrices, a smooth vector-valued object even when its norm vanishes. Conventional subsampling with $b=n^{0.7}$ undercovers badly in a weak-signal design, because of the missing finite-population factor and very small subsamples. We prove that fractional sampling ($b/n\to c\in(0,1)$) with the factor $\{bn/(n-b)\}^{1/2}$ is valid for independent data (Theorem~\ref{thm:subsample}). In the same design it restores coverage, partly through the finite-population factor and partly through larger subsets, but the resulting collection is conservative: it protects against excluding a sufficient set rather than discriminating among sets.
\item \emph{A latent Gaussian dependence model.} Units are functions of a latent Gaussian field with exponential (Mat\'ern-$\tfrac12$) spatial correlation and, for trajectories, Ornstein--Uhlenbeck (OU) temporal correlation, plus independent noise. Such data are $\rho$-mixing at an exponential rate uniformly in set size (Lemma~\ref{lem:mixing}), and the covariance of any two square-integrable scores is a power series $\sum_mc^mM_m$ in the latent correlation $c$ with positive semidefinite coefficients (Proposition~\ref{prop:mehler}). This gives the exact undercoverage of independence-based intervals.
\item \emph{Dependence-aware inference.} We prove a central limit theorem for augmented inverse-probability-weighted (AIPW) functional contrasts with parametric nuisances (Theorem~\ref{thm:depclt}). A parametric Mehler covariance estimator is consistent under finite chaos order, and a convex-chord estimator is asymptotically conservative for any chaos order and any overstatement of the spatial range (Theorem~\ref{thm:covest}). Block fractional sampling extends the collection calibration to dependent data (Theorem~\ref{thm:blockhalf}), under a linearization of the criterion tensor that we assume rather than prove; the selection procedure as a whole is therefore established for independent data only.
\item \emph{Evidence.} Simulations generate Poisson incidence trajectories from a Gaussian-process (GP)--OU latent field on lattices and on U.S.\ state geography. An analysis of state syphilis surveillance fits a Poisson GP--OU model, reports a mixed-model event study alongside two-way fixed effects, and compares dependence-aware and conventional intervals for the functional contrast.
\end{enumerate}

Section~\ref{sec:highdim} connects the criterion to recent high-dimensional multivariate theory \citep{umino2026automatic,egashira2024hierarchical,egashira2026kmeans,ishii2022geometric,bando2022consistency} through a deterministic stability bound, without claiming a high-dimensional inferential result. Two distinctions are essential throughout. An outer confidence collection protects against excluding a sufficient set; it does not certify every retained set, and at fixed error probability it does not support an oracle post-selection analysis. And Gaussian covariate laws within exposure groups are incompatible with nontrivial confounding under a uniform lower propensity bound, so we use strict positivity for identification and weighted conditions for estimation. The functional results concern outcomes isometric to a closed convex subset of a separable Hilbert space, such as univariate Wasserstein distributions, and complement Fr\'echet regression and distributional causal inference \citep{petersen2019frechet,lin2023causal} and scalar adjustment-set efficiency results \citep{hahn2004functional,rotnitzky2020efficient,henckel2022graphical}. The dependence theory assumes increasing-domain sampling and no interference. All proofs are in \ref{app:proofs}.

The causal framework is set out in Section~\ref{sec:ident}: unconfoundedness and a relative-overlap condition for general stochastic interventions, the average treatment effect and incremental-intervention effects for distribution-valued outcomes, and natural direct and indirect effects under sequential ignorability (Section~\ref{sec:mediation}). Every resulting estimator solves stacked estimating equations, so the dependent-data theory of Section~\ref{sec:dependence} covers all of them. Section~\ref{sec:geocausal} compares the approach, in theory and in application, with the point-process framework of \citet{papadogeorgou2022causal} and \citet{mukaigawara2025spatiotemporal}, implemented in the R package \texttt{geocausal}.

\section{Causal framework and identification}\label{sec:ident}
\subsection{Potential outcomes, interventions and causal assumptions}\label{sec:causal}
Let $O=(X,W,Y)$, where $X\in\R^p$, $W\in\mathcal W=\{1,\ldots,K\}$, and $Y$ takes values in a Polish space $\Omega$. The dimensions $p$ and $K$ are fixed. Each unit has potential outcomes $Y(w)$, $w\in\mathcal W$. Throughout we maintain the stable unit treatment value assumption \citep{rubin1980randomization}: consistency, $Y=Y(W)$, without hidden versions of the exposure, and no interference, so that a unit's potential outcomes do not depend on the exposures of other units. Section~\ref{sec:dependence} lets units be dependent but keeps no interference; Section~\ref{sec:geocausal} contrasts this with point-process designs in which spillover is itself the object of study.

Distribution-valued outcomes are averaged through an isometry $\iota$ from $\Omega$ onto a closed convex subset $\mathcal C$ of a separable Hilbert space $\Hcal$; write $\widetilde Y=\iota(Y)$ and $\widetilde Y(w)=\iota\{Y(w)\}$. If $\E\|\widetilde Y\|^2<\infty$, then $\E\widetilde Y\in\mathcal C$ corresponds to the Fr\'echet mean, because $\E\|\widetilde Y-a\|^2=\E\|\widetilde Y-\E\widetilde Y\|^2+\|a-\E\widetilde Y\|^2$. For one-dimensional Wasserstein outcomes, $\iota$ is the quantile embedding in $L^2(0,1)$.

An \emph{intervention} $F$ is a conditional law of the exposure given the covariates. Let $f$ denote the density of $F$ with respect to counting measure on $\mathcal W$; we let $f(w\mid x)$ depend on $x$ only through $x_S$ for a covariate set $S\subseteq[p]$. Setting $W=a$ is the deterministic intervention $f_a(w\mid x)=\I(w=a)$; a stochastic intervention spreads mass over several exposures \citep{munoz2012population,kennedy2019nonparametric,papadogeorgou2022causal}. The primary estimand is the \emph{average treatment effect} (ATE) of arm $a$ versus arm $b$,
\[
\Delta=\E\{\widetilde Y(a)-\widetilde Y(b)\}\in\Hcal,
\]
which contrasts the deterministic interventions $f_a$ and $f_b$. For Wasserstein outcomes it is the difference between the quantile functions of the Wasserstein barycentres of the two potential-outcome distributions \citep{lin2023causal}. Section~\ref{sec:estimands} places it among the stochastic-intervention effects. The observed assignment mechanism has density $e(w\mid x_S)=P(W=w\mid X_S=x_S)$, the propensity score.

We begin by introducing two causal assumptions. We first consider the unconfoundedness assumption, which ensures that the treatment assignment is random given the observed covariates \citep{papadogeorgou2022causal,bojinov2019time}. Formally, let $f$ denote the density of intervention $F$. The unconfoundedness assumption in our context is stated as follows.
\begin{assumption}[Unconfoundedness]\label{ass:unconf}
For every exposure $w$ with $P\{f(w\mid X_S)>0\}>0$ and every $v\in\mathcal W$,
\[
P\{W=v\mid X_S,\,Y(w)\}=e(v\mid X_S)\quad\text{almost surely; equivalently, }Y(w)\indep W\mid X_S .
\]
\end{assumption}
The second assumption replaces the bounded relative overlap condition $e(w\mid x)\ge c_0f(w\mid x)$, $c_0>0$, of \citet{papadogeorgou2022causal} by a moment condition. Let $m_w(x)=\E(\widetilde Y\mid X_S=x,W=w)$ and $\sigma_w^2(x)=\E\{\|\widetilde Y(w)-m_w(x)\|^2\mid X_S=x\}$, with the convention $f\sigma_w^2/e=0$ where $f(w\mid x)=0$.
\begin{assumption}[Relative overlap]\label{ass:overlap}
(a) $e(w\mid X_S)>0$ almost surely on the event $\{f(w\mid X_S)>0\}$. (b) $\E\bigl\{\sum_wf(w\mid X_S)^2\sigma_w^2(X_S)/e(w\mid X_S)\bigr\}<\infty$.
\end{assumption}
Part~(b) holds under bounded relative overlap whenever $\E\sigma_w^2(X_S)<\infty$, but it does not require the ratio $f/e$ to be bounded. This matters for the ATE: for the deterministic interventions $f_a$ and $f_b$, a bounded ratio $f/e$ means uniform overlap, $e(a\mid X_S)\ge c_0$ and $e(b\mid X_S)\ge c_0$, which the Gaussian covariate laws of Proposition~\ref{prop:overlap} exclude unless $W\indep X$. Part~(b) instead requires only $\E\{\sigma_a^2(X_S)/e(a\mid X_S)+\sigma_b^2(X_S)/e(b\mid X_S)\}<\infty$.

\subsection{Causal estimands and their identification}\label{sec:estimands}
For an intervention $F$ define the interventional mean
\[
\mu(F)=\E\Bigl\{\sum_{w}f(w\mid X_S)\,\widetilde Y(w)\Bigr\}\in\mathcal C,
\]
the Fr\'echet mean of the potential outcome $Y(W^F)$ when $W^F$ is drawn from $F(\cdot\mid X)$ independently of the potential outcomes given $X$. Since $\mu(f_a)=\E\widetilde Y(a)$, the ATE of Section~\ref{sec:causal} is the contrast $\Delta=\mu(f_a)-\mu(f_b)$ of two deterministic interventions. More generally, a stochastic-intervention effect is $\tau(F',F'')=\mu(F')-\mu(F'')$. For binary exposure with $e(x)=e(1\mid x)$, the incremental intervention $F_\delta$ of \citet{kennedy2019nonparametric} multiplies each unit's odds of exposure by $\delta>0$,
\begin{equation}\label{eq:incremental}
f_\delta(1\mid x)=\frac{\delta e(x)}{\delta e(x)+1-e(x)},
\end{equation}
and $\tau_\delta=\mu(F_\delta)-\E\widetilde Y$ compares it with the observed assignment, which is the case $\delta=1$. Binary exposures are labelled $\mathcal W=\{0,1\}$. Mediation estimands are defined in Section~\ref{sec:mediation}.
\begin{theorem}[Identification of interventional means]\label{thm:causalid}
Suppose Assumptions~\ref{ass:unconf} and~\ref{ass:overlap}(a) hold and $\E\|\widetilde Y(w)\|<\infty$ for every $w$ charged by $F$. Then
\begin{equation}\label{eq:gformula}
\mu(F)=\E\Bigl\{\sum_wf(w\mid X_S)\,m_w(X_S)\Bigr\}=\E\Bigl\{\frac{f(W\mid X_S)}{e(W\mid X_S)}\,\widetilde Y\Bigr\}.
\end{equation}
Under Assumption~\ref{ass:overlap}(b) and $\E\|\widetilde Y(w)\|^2<\infty$, the score
\[
\psi_F=\sum_wf(w\mid X_S)\,m_w(X_S)+\frac{f(W\mid X_S)}{e(W\mid X_S)}\{\widetilde Y-m_W(X_S)\}
\]
is square integrable with $\E\psi_F=\mu(F)$. If $f$ does not depend on $e$ and $\widetilde\psi_F$ is $\psi_F$ with $(m_w,e)$ replaced by working functions $(\widetilde m_w,\widetilde e)$, where $\widetilde e>0$ wherever $f>0$ and both terms of $\widetilde\psi_F$ are integrable, then $\E\widetilde\psi_F=\mu(F)$ whenever $\widetilde m_w=m_w$ for all $w$ or $\widetilde e=e$. For the incremental intervention \eqref{eq:incremental}, $e(w\mid x)\ge\min(\delta,\delta^{-1})f_\delta(w\mid x)$ for every propensity score. Assumption~\ref{ass:overlap} then holds whenever $\E\sigma_w^2(X_S)<\infty$, even if $e(X_S)$ is not bounded away from zero and one.
\end{theorem}
When $f$ does not depend on the law of the data, $\psi_F-\mu(F)$ is the nonparametric efficient influence function \citep{munoz2012population}. For the incremental intervention $f_\delta$ depends on $e$, and the efficient influence function gains a term for estimating $e$ \citep{kennedy2019nonparametric}; it appears in Section~\ref{sec:stochastic}. The last statement of Theorem~\ref{thm:causalid} is the reason incremental effects are useful here: they are identified without positivity, including under the Gaussian confounding of Proposition~\ref{prop:overlap}.

\paragraph{Unconfoundedness with spatial data} \citet{papadogeorgou2022causal} require the treatment pattern of each period to be independent of all past and future potential outcomes given the observed history, because with spillover and carryover every pattern can affect every later outcome; \citet{bojinov2019time} impose the analogous non-anticipation condition on assignment paths in time-series experiments. Without interference a unit-level statement suffices. Under the latent model of Section~\ref{sec:dependence}, $W_i$ may be correlated with neighbours' potential outcomes $Y_k(w)$ through the latent field; such exposure clustering affects variances but not identification. What Assumption~\ref{ass:unconf} excludes is a latent component that affects both the exposure and the outcome of the same unit and is not captured by $X_S$, that is, unmeasured spatial confounding \citep{paciorek2010importance,reich2021spatial}. A sufficient condition, in the notation of Assumption~\ref{ass:latent}, is that $Y_i(w)$ depends on $(\Gamma_i,\varepsilon_i)$ only through $X_{iS}$ and through latent and noise components that are independent of $(X_{iS},W_i)$. The designs of Section~\ref{sec:numerical} are of this form.

\subsection{Sufficient adjustment sets}\label{sec:sufficient}
A set $A\subseteq[p]$ is sufficient for arm $w$ if Assumption~\ref{ass:unconf} holds with $S=A$ for the intervention $f_w$, that is, $Y(w)\indep W\mid X_A$; write $\Acal_w$ for these sets. Theorem~\ref{thm:causalid} applies to every $S\in\bigcap_{w}\Acal_w$, the intersection over the arms charged by $F$. Sufficiency is not assumed to be upward closed: adding a covariate can open a collider path.

\begin{assumption}[Selection design]\label{ass:ident}
For every $w$, $[p]\in\Acal_w$ and $\pi_w(X)=P(W=w\mid X)>0$ almost surely. Every arm has positive marginal probability. Conditional on $W=v$, $X\sim N(\nu_v,\Sigma_v)$ with $\Sigma_v\succ0$. The central subspaces for $Y(w)\mid X$ and $W\mid X$ exist, with orthonormal bases $B_w$ and $B_W$. For each $w,A$, if $B_w\tp X$ and $B_W\tp X$ are conditionally dependent given $X_A$, there are bounded measurable $h,k$ such that
\[
\Cov\bigl[\E\{h(Y(w))\mid B_w\tp X\},\E\{k(W)\mid B_W\tp X\}\mid X_A\bigr]\ne0
\]
on a set of positive probability.
\end{assumption}
The last condition is a detectability assumption: it rules out cancellations between the outcome and exposure mechanisms. It is not implied by Gaussian covariates. The central subspace is the smallest linear subspace making the response conditionally independent of $X$ given its projection \citep{cook1998regression}. Its dimension can be zero, and the dimension for a categorical response need not be at most $K-1$.

For $A\subseteq[p]$, define the residual covariance matrix, including zero rows and columns on $A$,
\[
H_v(A)=\Sigma_v-\Sigma_{v,\cdot A}\Sigma_{v,AA}^{-1}\Sigma_{v,A\cdot},\qquad
H_v(\varnothing)=\Sigma_v,\quad H_v([p])=0.
\]
Let $P_w=B_wB_w\tp$, $P_W=B_WB_W\tp$, and
\begin{equation}\label{eq:criterion}
T_{wvA}=P_wH_v(A)P_W,\qquad Q_w(A)=\sum_{v=1}^K\|T_{wvA}\|_{\rm op}.
\end{equation}
The identity $\|T_{wvA}\|_{\rm op}=\|B_w\tp H_v(A)B_W\|_{\rm op}$ follows because multiplication by orthonormal bases preserves nonzero singular values. Thus \eqref{eq:criterion} retains the original criterion while avoiding arbitrary basis orientations.

\begin{theorem}[Finite-exposure identification of sufficient sets]\label{thm:ident}
Under Assumption~\ref{ass:ident}, $Q_w(A)=0$ if and only if $A\in\Acal_w$. The observed conditional law of $(X,Y)$ given $W=w$ identifies the same outcome central subspace as $Y(w)\mid X$.
\end{theorem}
This is a finite-exposure and object-response extension of the binary criterion of \citet{luo2024exhaustive}. The proof uses mutual absolute continuity and remains valid when the reduced predictors overlap or have degenerate joint covariance.

\begin{proposition}[Gaussian overlap incompatibility]\label{prop:overlap}
Under the Gaussian design in Assumption~\ref{ass:ident}, if $\pi_v(x)\ge\eta>0$ for every $v$ almost surely, all $(\nu_v,\Sigma_v)$ coincide and $W\indep X$. In contrast, if $\Sigma_v=\Sigma$ but the means differ, strict positivity holds, $W$ may depend on $X$, and $\E\{\pi_v(X)^{-r}\}<\infty$ for every finite $r>0$.
\end{proposition}
Thus uniform overlap is not imposed in the identification theorem. The equal-covariance design also makes the logistic propensity model exactly correct, a fact used in Section~\ref{sec:dependence}.

\paragraph{Kernels and coordinate transformations}
For object outcomes, use a fixed bounded continuous characteristic kernel $\kappa$ and its feature distance
$\rho(y,y')=\{\kappa(y,y)+\kappa(y',y')-2\kappa(y,y')\}^{1/2}$.
\begin{proposition}[Information preservation]\label{prop:kernel}
The feature map of such a kernel preserves the response central subspace. If $\E\|X\|^\alpha<\infty$, $0<\alpha<2$, distance covariance computed with $\|x-x'\|^\alpha$ and $\rho$ is zero exactly under independence.
\end{proposition}
Proposition~\ref{prop:kernel} concerns information preservation and independence detection; recovery of the central subspace by a dimension-constrained maximizer is a separate assumption below. A known invertible coordinatewise transformation preserves sufficiency, but the Gaussian design must then hold in the transformed coordinates; pooled empirical normal scores are not covered.

\section{Subspace estimation and collection inference}\label{sec:selection}
\subsection{A \texorpdfstring{root-$n$}{root-n} expansion with estimated whitening}
Let $V$ denote either an outcome or the categorical exposure. Write $Z=\Sigma_X^{-1/2}(X-\mu)$ and $a_D(z,z')=\|D\tp(z-z')\|^\alpha$, $0<\alpha<2$. The response distance is either $\rho(v,v')=\|v-v'\|^\alpha$, a bounded feature distance, or $\I(v\ne v')$. Define
\begin{equation}\label{eq:population}
\Gamma(D)=\E(a_{12}\rho_{12})+\E(a_{12})\E(\rho_{34})-2\E(a_{12}\rho_{13}),
\end{equation}
where $a_{ij}=a_D(Z_i,Z_j)$. Let $h_D$ be the symmetrization of the integrand in \eqref{eq:population} over four independent observations. Its order-four U-statistic is $\Gamma_n(D)$. Equivalently,
\[
\Gamma_n(D)=\frac{1}{n(n-3)}\sum_{i\ne j}\widetilde a_{ij}\widetilde\rho_{ij},
\]
where the tildes denote U-centering \citep{szekely2014partial}: for a zero-diagonal distance matrix $D$, its U-centered off-diagonal entries are
$D_{ij}-(D_{i+}+D_{+j})/(n-2)+D_{++}/\{(n-1)(n-2)\}$; the diagonal remains zero.

Let $C_0$ span the standardized central subspace of dimension $d$, and use the Grassmann chart
\[
C(\theta)=(C_0+C_{0\perp}\theta)(I+\theta\tp\theta)^{-1/2},
\quad \theta\in\R^{(p-d)\times d}.
\]
Set $R=\widehat\Sigma_X^{-1/2}\Sigma_X^{1/2}$, $G_n(\theta,R)=\Gamma_n\{R\tp C(\theta)\}$, and
$\tau(o;\theta,R)=\E h_{R\tp C(\theta)}(o,O_2,O_3,O_4)$.

\begin{assumption}[SDR regularity]\label{ass:sdr}
Observations are i.i.d., $\Sigma_X\succ0$, and $0<d<p$ is known. The only population maximizers on the Stiefel manifold are $C_0O$ for orthogonal $O$. We have
$\E\|X\|^4<\infty$, $\E(1+\|X\|)^{4\alpha}<\infty$, and $\E\rho(V,v_0)^4<\infty$.
On a compact neighborhood of $(0,I)$, $\tau(o;\theta,R)$ is twice continuously differentiable for almost every $o$, with its first and second derivatives bounded by $M(o)$ satisfying $\E M(O)^2<\infty$. The Hessian $\mathbb V=\partial^2_\theta G(0,I)$ is negative definite.
The computed maximizer of the empirical objective is within $o_P(n^{-1})$ of its global maximum in objective value.
\end{assumption}
The averaged smoothness condition is substantive, especially for $\alpha<1$: Gaussian predictor density alone does not control the response-weighted conditional integrals. The theorem is conditional on this assumption. For $d=0$ or $d=p$, the subspace projector is fixed. Dimension estimation requires an additional consistent procedure and is not supplied by the present results.

\begin{theorem}[Projector expansion]\label{thm:rate}
Under Assumption~\ref{ass:sdr}, for the U-statistic maximizer based on $\widehat Z_i=\widehat\Sigma_X^{-1/2}(X_i-\overline X)$,
\begin{equation}\label{eq:bahadur}
\sqrt n\,\widehat\theta=-\mathbb V^{-1}\frac1{\sqrt n}\sum_{i=1}^n\xi(O_i)+o_P(1),
\end{equation}
where
\[
\xi(o)=4\partial_\theta\tau(o;0,I)+\mathbb K\mathcal L_\Sigma[(x-\mu)(x-\mu)\tp-\Sigma_X],
\]
$\mathbb K=\partial_R\partial_\theta G(0,I)$ and $\mathcal L_\Sigma$ is the derivative of $S\mapsto S^{-1/2}\Sigma_X^{1/2}$ at $S=\Sigma_X$. The projector of $\widehat B=\widehat\Sigma_X^{-1/2}\widehat C$ is root-$n$ consistent and asymptotically linear. Joint expansions hold for finitely many armwise fits and the exposure fit when their respective regularity conditions hold.
\end{theorem}
For an armwise fit the influence function is multiplied by $\I(W=w)/P(W=w)$; the corresponding conditional covariance estimate has the analogous weighting. Joint convergence therefore follows from one multivariate i.i.d. CLT, not from independence of the fits.

The implemented predictor distance is $(\|D\tp(z-z')\|^2+\epsilon_n)^{\alpha/2}$ off the diagonal, with $\epsilon_n=n^{-4/\alpha}$; its uniform objective discrepancy is $O_P(n^{-2})$ (\ref{app:rate}). Multistart gradient ascent remains a local optimization method: the global objective-gap condition is not certified by a small gradient norm. Fixed kernel bandwidths are used in all experiments.

\subsection{Consistent point selection}
Let $\Fcal$ be a fixed finite family of candidate sets containing $[p]$, so at least one candidate is sufficient for every arm. Define $\Acal_w^\Fcal=\Acal_w\cap\Fcal$ and estimate \eqref{eq:criterion}. Order the $M=|\Fcal|$ values decreasingly. For $1\le k<M$, put
\begin{equation}\label{eq:ridge}
R_w(k)=\frac{\widehat Q_w(A_{(k+1)})+c_n}{\widehat Q_w(A_{(k)})+c_n},\quad R_w(0)=c_0\in(0,1),\quad
c_n=c_*n^{-1/2}\log n\max\{\max_A\widehat Q_w(A),\kappa_0\},
\end{equation}
where $c_*,\kappa_0>0$. Retain indices $k>\widehat k_w$, where $\widehat k_w$ minimizes $R_w$ over $0,\ldots,M-1$.
\begin{theorem}[Ridge recovery]\label{thm:ridge}
Under Theorem~\ref{thm:ident}, the joint expansions in Theorem~\ref{thm:rate}, and root-$n$ consistent arm covariance estimates, the ridge collection equals $\Acal_w^\Fcal$ simultaneously for all $w$ with probability tending to one. This includes the case where every candidate set is sufficient.
\end{theorem}
The statement is pointwise for fixed population criteria, not uniform when an insufficient set approaches the null at order $n^{-1/2}$.

\subsection{A simultaneous outer collection: subsampling and fractional sampling}\label{sec:collection}
Collect all entries of $T_{wvA}$ into a finite vector $\mathbf T$. This vector is a smooth function of projectors and positive-definite covariance matrices; no norm is differentiated at zero. Under the conditions of Theorem~\ref{thm:ridge} it is asymptotically linear,
\begin{equation}\label{eq:Tlinear}
\widehat{\mathbf T}_n-\mathbf T=\frac1n\sum_{i=1}^n\psi_T(O_i)+o_P(n^{-1/2}),
\end{equation}
with $\E\psi_T=0$ and $\E\|\psi_T\|^2<\infty$, by the delta method applied to Theorem~\ref{thm:rate} and the covariance expansions. Set
$\mathcal N(t)=\max_{w,A}\sum_v\|t_{wvA}\|_{\rm op}$.
Take simple random subsets $I$ of size $b$ without replacement, recompute the entire estimator to obtain $\widehat{\mathbf T}_I$, and form
\begin{equation}\label{eq:ZI}
Z_I=s_{n,b}\,\mathcal N(\widehat{\mathbf T}_I-\widehat{\mathbf T}_n),\qquad s_{n,b}=\Bigl(\frac{bn}{n-b}\Bigr)^{1/2}.
\end{equation}
Let $q_{n,b}(1-\gamma)$ be the conditional quantile of $Z_I$ and retain
\begin{equation}\label{eq:subsample}
\widehat\Acal_w^{\rm out}=\{A\in\Fcal:\sqrt n\widehat Q_w(A)\le q_{n,b}(1-\gamma)\}.
\end{equation}
For $b/n\to0$, $s_{n,b}\sim\sqrt b$ and \eqref{eq:ZI} is ordinary subsampling \citep{politis1994subsamples}. For $b/n\to c\in(0,1)$ it is the delete-$(n-b)$ jackknife histogram of \citet{wu1990jackknife}; we call it fractional sampling and use $c=1/2$.

\begin{theorem}[Simultaneous outer coverage]\label{thm:subsample}
Suppose the conditions of Theorem~\ref{thm:ridge} hold, so that \eqref{eq:Tlinear} holds, and that the optimization condition of Assumption~\ref{ass:sdr} holds for the fits computed on subsets as well as on the full sample. Let $Z\sim N\{0,\Var\psi_T(O)\}$ and suppose the distribution of $\mathcal N(Z)$ is continuous and strictly increasing at its $(1-\gamma)$ quantile. Let either (i) $b\to\infty$ and $b/n\to0$, or (ii) $b/n\to c\in(0,1)$. Then $q_{n,b}(1-\gamma)\to q(1-\gamma)$, the $(1-\gamma)$ quantile of $\mathcal N(Z)$, in probability, and
\begin{align*}
\liminf_n P\{\Acal_w^\Fcal\subseteq\widehat\Acal_w^{\rm out}\text{ for every }w\}&\ge1-\gamma,\\
P\{\widehat\Acal_w^{\rm out}\subseteq\Acal_w^\Fcal\text{ for every }w\}&\to1.
\end{align*}
The result also holds using $B\to\infty$ independently drawn subsets conditional on the data. Neither case assumes ordinary-bootstrap consistency.
\end{theorem}

Case (i) uses only the weak limit of $\widehat{\mathbf T}_n$; case (ii) uses the linear representation \eqref{eq:Tlinear} and a finite-population central limit theorem \citep{hajek1960limiting}. Case (ii) is preferable in moderate samples. With overlapping subsets $\Var(\widehat{\mathbf T}_I-\widehat{\mathbf T}_n)\approx(1/b-1/n)\Var\psi_T$, so scaling by $\sqrt b$ shrinks the calibrating distribution by $(1-b/n)^{1/2}$ (0.89 at $n=200$ for $b=n^{0.7}$). Moreover, projector errors saturate at very small $b$ instead of growing like $b^{-1/2}$. Both effects make subsampling anticonservative, and fractional sampling avoids the first exactly and the second largely. The second assertion excludes fixed insufficient sets; it does not force inclusion of all sufficient sets with probability tending to one, and choosing a set within an outer collection is not covered by the oracle results below unless its selection consistency is established separately.\subsection{Connection with high-dimensional multivariate theory}\label{sec:highdim}
Since $B_w\tp H_v(A)B_W=\Cov(B_w\tp R_{v,A},B_W\tp R_{v,A}\mid W=v)$ with $R_{v,A}=X-\E(X\mid X_A,W=v)$, the criterion is a residual cross-covariance, so high-dimensional covariance estimators \citep{umino2026automatic} are natural candidates. The following elementary bound, in which $p$, $K$ and the candidate family may vary with $n$, states what such an estimator would have to deliver.
\begin{proposition}[Transfer of uniform matrix error]\label{prop:transfer}
Let the true and estimated $P_w,P_W$ be orthogonal projectors. Suppose uniformly over all relevant indices that
\[
\|H_v(A)\|_{\rm op}\le L_n,\quad
\|\widehat H_v(A)-H_v(A)\|_{\rm op}\le h_n,\quad
\|\widehat P_w-P_w\|_{\rm op}\le e_{Y,n},\quad
\|\widehat P_W-P_W\|_{\rm op}\le e_{W,n}.
\]
Then, for $\widehat T_{wvA}=\widehat P_w\widehat H_v(A)\widehat P_W$ and its corresponding $\widehat Q_w(A)$,
\begin{equation}\label{eq:transfer}
\max_{w,A}|\widehat Q_w(A)-Q_w(A)|
\le K_n\{h_n+L_n(e_{Y,n}+e_{W,n})\}.
\end{equation}
If the right side is $O_P(r_n)$, the zero sets are exactly the sufficient candidates, and their smallest positive criterion is $\delta_n$, a deterministic threshold $t_n$ satisfying $r_n=o(t_n)$ and $t_n=o(\delta_n)$ recovers all candidate sufficiency labels with probability tending to one. If no positive criteria exist, only $r_n=o(t_n)$ is needed.
\end{proposition}
Proposition~\ref{prop:transfer} is a point-selection implication, not a growing-dimensional inferential result; all collection results remain fixed dimensional. Section~S5 of the Supplementary Material discusses the connections with high-dimension, low-sample-size clustering theory \citep{egashira2024hierarchical,egashira2026kmeans,ishii2022geometric,bando2022consistency}.

\section{Average treatment effects, stochastic interventions and mediation}\label{sec:effects}
\subsection{Average treatment effects under nonuniform overlap}\label{sec:ate}
Fix distinct arms $a,b$ and a common sufficient set $S\in\Acal_a\cap\Acal_b$, so that the ATE $\Delta$ of Section~\ref{sec:causal} is identified by Theorem~\ref{thm:causalid}. Define
\begin{align*}
m_v(x)&=\E\{\widetilde Y(v)\mid X_S=x\}=\E(\widetilde Y\mid X_S=x,W=v),\\
\Delta&=\E\{\widetilde Y(a)-\widetilde Y(b)\},\qquad \tau_S=m_a-m_b,\\
\Delta_{C,n}&=n^{-1}\sum_i\tau_S(X_{iS}),\qquad
\Delta_{F,n}=n^{-1}\sum_i\{\widetilde Y_i(a)-\widetilde Y_i(b)\}.
\end{align*}
These are the population contrast, the conditional sample average contrast, and the contrast of sample potential outcomes. Differences of quantile functions lie in the ambient Hilbert space and are not themselves necessarily quantile functions; their $L^2$ loss should not be called a Wasserstein distance between distributions.

For $\pi_v(x)=e(v\mid x)=P(W=v\mid X_S=x)$ define
\[
\psi_v=m_v(X_S)+\frac{\I(W=v)}{\pi_v(X_S)}\{\widetilde Y-m_v(X_S)\},\quad
\varphi=\psi_a-\psi_b-\Delta.
\]
Estimate $\Delta$ by the average of cross-fitted scores $\widehat\psi_a-\widehat\psi_b$, using a fixed number of folds. The following weighted conditions accommodate propensities approaching zero.

\begin{assumption}[Weighted nuisance control]\label{ass:nuisance}
Potential outcomes have finite second moments. For $v=a,b$, $\pi_v>0$, fitted propensities are positive, and
$\E\{\sigma_v^2(X_S)/\pi_v(X_S)\}<\infty$, where
$\sigma_v^2(x)=\E\{\|\widetilde Y(v)-m_v(x)\|^2\mid X_S=x\}$.
On each training fold, writing $d_m=\widehat m_v-m_v$ and $d_\pi=\widehat\pi_v-\pi_v$, integration over a fresh $X_S$ satisfies
\begin{align}
\E\{(1+\pi_v/\widehat\pi_v^2)\|d_m\|^2\}&=o_P(1),\label{eq:weighted1}\\
\E\{\sigma_v^2 d_\pi^2/(\pi_v\widehat\pi_v^2)\}&=o_P(1),\label{eq:weighted2}\\
\E\{\|d_m\||d_\pi|/\widehat\pi_v\}&=o_P(n^{-1/2}).\label{eq:weighted3}
\end{align}
For inference on $\Delta_{F,n}$, require joint sufficiency $\{Y(a),Y(b)\}\indep W\mid X_S$.
\end{assumption}
Bounded inverse propensities, bounded residual variance, $L^2$ nuisance consistency and an $o_P(n^{-1/2})$ product rate suffice, but that special case is incompatible with distinct Gaussian arm distributions (Proposition~\ref{prop:overlap}). Clipping is a numerical stabilization, not part of identification. If the propensity is replaced by $e_c=\min\{\max(\pi_v,c),1-c\}$ while the outcome regression $\widetilde m_v$ is misspecified, the arm-$v$ mean has bias $\E[\{1-\pi_v(X_S)/e_c(X_S)\}\{\widetilde m_v(X_S)-m_v(X_S)\}]$, so the propensity-correct branch of double robustness is lost wherever $\pi_v\notin[c,1-c]$. Clipped AIPW therefore estimates the ATE consistently in two cases, under the regularity conditions of Theorem~\ref{thm:effects}: if the outcome regression is correct, with suitable integrability; or if the propensity estimator is consistent and clipping is asymptotically inactive, that is, no true propensity lies outside $[c,1-c]$. True propensities inside $[c,1-c]$ do not by themselves protect against a misspecified propensity model. Otherwise it estimates a clipped functional, and the application reports such estimates as clipped AIPW associations with a sensitivity analysis.

\begin{theorem}[Functional limits and covariance ordering]\label{thm:effects}
Under i.i.d. sampling, sufficiency of the fixed common set $S$ and Assumption~\ref{ass:nuisance},
\[
\widehat\Delta-\Delta=\Pn\varphi+o_P(n^{-1/2}).
\]
The root-$n$ errors centered at $\Delta,\Delta_{C,n},\Delta_{F,n}$ converge in $\Hcal$ to centered Gaussian elements with covariance operators $\mathcal C_P,\mathcal C_C,\mathcal C_F$, respectively. With $\varepsilon_v=\widetilde Y(v)-m_v(X_S)$,
\begin{align*}
\mathcal C_C&=\E\{\Cov(\varepsilon_a\mid X_S)/\pi_a+\Cov(\varepsilon_b\mid X_S)/\pi_b\},\\
\mathcal C_P&=\mathcal C_C+\Cov\{\tau_S(X_S)\},\\
\mathcal C_F&=\mathcal C_C-\E\Cov(\varepsilon_a-\varepsilon_b\mid X_S).
\end{align*}
Consequently $\mathcal C_F\preceq\mathcal C_C\preceq\mathcal C_P$. The same limits hold for a selected set equaling a fixed sufficient set with probability tending to one.
\end{theorem}
These are joint-sampling limits with random centers for the last two estimands. The empirical covariance of cross-fitted scores estimates $\mathcal C_P$ consistently in trace norm and is conservative for the other two centers; the cross-potential-outcome covariance in $\mathcal C_F$ is generally unidentified.

\begin{theorem}[Ordering nested sufficient sets]\label{thm:efficiency}
Let $S\subset S'$ both be sufficient for arms $a,b$, and assume all covariance operators below are finite. If $X_{S'\setminus S}\indep W\mid X_S$, then $\mathcal C_{S'}\preceq\mathcal C_S$. If instead $X_{S'\setminus S}\indep\{Y(a),Y(b)\}\mid X_S$, then $\mathcal C_S\preceq\mathcal C_{S'}$.
For a finite nonempty family of sufficient sets recovered consistently, uniformly consistent trace estimates select its unique trace minimizer with probability tending to one.
\end{theorem}
The ordering concerns nonparametric AIPW influence functions; a correctly specified parametric g-computation model can have a different efficiency bound, so no dominance over every full-adjustment estimator is claimed.

\paragraph{Conditional targets and shape}
For nested sufficient sets, $\tau_S(X_S)=\E\{\tau_{S'}(X_{S'})\mid X_S\}$, so the conditional target changes when information is coarsened. If $g$ is a scalar summary, sufficiency for $Y(w)$ implies sufficiency for $g\{Y(w)\}$, but the converse can fail. Consider $P(W=1)=1/2$, $X\mid W=w\sim N(w\mu,1)$ with $\mu>0$, and $Y(0)=Y(1)$ equal to the uniform distribution on $[-e^X,e^X]$. This defines zero causal effect with shape confounding. Its mean is always zero, whereas its unadjusted quantile contrast is
$(2u-1)\{\E(e^X\mid W=1)-\E(e^X\mid W=0)\}\ne0$ for $u\ne1/2$.

\subsection{Stochastic and incremental interventions}\label{sec:stochastic}
For an intervention $F$ such that $S$ is sufficient for every arm it charges, $\widehat\mu(F)$ averages cross-fitted scores $\widehat\psi_F$ of Theorem~\ref{thm:causalid}. When $f$ is known, the proof of Theorem~\ref{thm:effects} applies after replacing $\I(W=v)/\pi_v$ by $f(W\mid X_S)/e(W\mid X_S)$. Under Assumption~\ref{ass:overlap}(b) and the analogues of \eqref{eq:weighted1}--\eqref{eq:weighted3} with these weights, $\sqrt n\{\widehat\mu(F)-\mu(F)\}$ converges in $\Hcal$ to a centred Gaussian element with covariance operator $\Cov\psi_F$. For the incremental intervention \eqref{eq:incremental}, with $\psi_v$ as in Section~\ref{sec:ate}, the uncentred efficient influence function of $\mu(F_\delta)$ is \citep{kennedy2019nonparametric}
\begin{equation}\label{eq:incif}
\varphi_\delta=\frac{\delta e(X_S)\psi_1+\{1-e(X_S)\}\psi_0}{\delta e(X_S)+1-e(X_S)}+\frac{\delta\{m_1(X_S)-m_0(X_S)\}\{W-e(X_S)\}}{\{\delta e(X_S)+1-e(X_S)\}^2}.
\end{equation}
Since $e\psi_1=em_1+\I(W=1)(\widetilde Y-m_1)$, every weight in \eqref{eq:incif} is bounded by $\max(\delta,\delta^{-1})$, so neither clipping nor positivity is required. If the propensity model is correct, $\E\varphi_\delta=\mu(F_\delta)$ whatever the outcome regressions, because then $\E(\widetilde\psi_v\mid X_S)=m_v(X_S)$ for working regressions $\widetilde m_v$ and $\E\{W-e(X_S)\mid X_S\}=0$. A correct outcome model alone does not suffice, because $F_\delta$ is defined through $e$ and the remainder involves the propensity error \citep{kennedy2019nonparametric}. With parametric nuisances, the propensity, outcome and target equations are stacked as in Section~\ref{sec:dependence}.

\subsection{Causal mediation analysis}\label{sec:mediation}
Let $W\in\{0,1\}$ and let $M$ be a mediator taking values in a Polish space $\mathcal M$, observed after the exposure and before the outcome. Potential mediators $M(w)$ and potential outcomes $Y(w,m)$ satisfy consistency, $M=M(W)$ and $Y=Y(W,M)$, and composition, $Y(w)=Y\{w,M(w)\}$. For $a,b\in\{0,1\}$ let $\theta_{ab}=\E\,\widetilde Y\{a,M(b)\}\in\mathcal C$. The natural direct and indirect effects \citep{robins1992identifiability,pearl2001direct} are
\[
\mathrm{NDE}=\theta_{10}-\theta_{00},\qquad \mathrm{NIE}=\theta_{11}-\theta_{10},\qquad \mathrm{NDE}+\mathrm{NIE}=\Delta ,
\]
an exact decomposition of the ATE in $\Hcal$. For a level--timing outcome, the NIE separates the part of the change in amount and timing of incidence that runs through the mediator.
\begin{assumption}[Sequential ignorability]\label{ass:seqign}
For $a,b\in\{0,1\}$ and every $m\in\mathcal M$: (i) $\{Y(a,m),M(b)\}\indep W\mid X_S$; (ii) $Y(a,m)\indep M(b)\mid W=b,X_S$; (iii) $e(w\mid X_S)>0$ for $w=0,1$, and the conditional law $P_{M\mid x,b}$ of $M$ given $X_S=x,W=b$ is absolutely continuous with respect to $P_{M\mid x,a}$ for almost every $x$, with density $r_{ab}(m,x)$.
\end{assumption}
Part~(i) is Assumption~\ref{ass:unconf} for the pair $\{Y(a,m),M(b)\}$; part~(ii) excludes unmeasured mediator--outcome confounding and mediator--outcome confounders affected by the exposure \citep{imai2010general}. It is a cross-world condition that no design can guarantee. Let $\mu_a(x,m)=\E(\widetilde Y\mid X_S=x,W=a,M=m)$ and $\eta_{ab}(x)=\int\mu_a(x,m)\,P_{M\mid x,b}(dm)$.
\begin{theorem}[Identification and multiply robust estimation of mediation effects]\label{thm:mediation}
Suppose Assumption~\ref{ass:seqign} holds and $\E\int\E\{\|\widetilde Y(a,m)\|\mid X_S\}\,P_{M\mid X_S,b}(dm)<\infty$, which allows outcome regressions that are unbounded in $m$, such as linear ones. Then:
\begin{enumerate}[label=(\alph*),leftmargin=1.6em,itemsep=1pt]
\item $\theta_{ab}=\E\,\eta_{ab}(X_S)$, and $r_{ab}(m,x)=\{e(a\mid x)P(W=b\mid X_S=x,M=m)\}/\{e(b\mid x)P(W=a\mid X_S=x,M=m)\}$.
\item Let
\[
\psi_{ab}=\frac{\I(W=a)\,r_{ab}(M,X_S)}{e(a\mid X_S)}\{\widetilde Y-\mu_a(X_S,M)\}+\frac{\I(W=b)}{e(b\mid X_S)}\{\mu_a(X_S,M)-\eta_{ab}(X_S)\}+\eta_{ab}(X_S),
\]
and let $\widetilde\psi_{ab}$ be $\psi_{ab}$ evaluated at working models $(\widetilde e,\widetilde P_M,\widetilde\mu_a)$ for the propensity, the mediator law and the outcome regression, with $\widetilde r_{ab}$ and $\widetilde\eta_{ab}$ computed from them. If $\E\|\widetilde\psi_{ab}\|<\infty$, then $\E\widetilde\psi_{ab}=\theta_{ab}$ whenever at least two of the three working models are correct.
\item If $\E\|\psi_{ab}\|^2<\infty$, then $\psi_{ab}-\theta_{ab}$ is the nonparametric efficient influence function of the identifying functional $\E\,\eta_{ab}(X_S)$, and with $a=b$ it reduces to the AIPW score $\psi_a$ of Section~\ref{sec:ate}.
\end{enumerate}
\end{theorem}
Part~(a) is the mediation formula \citep{pearl2001direct,imai2010general}, here for Hilbert-valued outcomes. Parts~(b) and (c) extend the triply robust estimator of \citet{tchetgen2012semiparametric} to such outcomes. Replacing $M(b)$ in $\theta_{ab}$ by a random draw from its conditional law given $X_S$ defines interventional direct and indirect effects \citep{didelez2006direct,vanderweele2014effect}. Without exposure-induced mediator--outcome confounders they are identified by the same formula without the cross-world part of (ii). With a measured exposure-induced confounder they remain identified by a modified g-formula that also integrates over that confounder, whereas natural effects are not identified \citep{vanderweele2014effect}. The mediation effects of \citet{mukaigawara2025spatiotemporal}, defined through joint stochastic interventions on the treatment and mediator distributions, are of this interventional type. Two features of distribution-valued outcomes are worth noting. First, the timing component of the level--timing outcome of Section~\ref{sec:application} is invariant to a multiplicative ascertainment factor that is constant over the window. A mediator that acts only through such ascertainment therefore has no indirect effect on timing, and a nonzero timing NIE indicates time-varying ascertainment or a genuine change in the course of incidence. Second, the theory of Section~\ref{sec:dependence} applies to the stacked parametric version of $\psi_{ab}$, which is used in Sections~\ref{sec:sim-mediation} and~\ref{sec:app-causal}.

\section{Latent Gaussian spatio-temporal dependence}\label{sec:dependence}
Surveillance units are not independent. Neighbouring jurisdictions share epidemics, policies cluster geographically, and incidence is persistent. We model this dependence through a latent Gaussian field, the standard structure of disease-mapping models \citep{besag1991bayesian,rue2009approximate,banerjee2014hierarchical}, and derive consequences for inference without assuming that the observed variables are Gaussian.

\subsection{The model and its mixing properties}
\begin{assumption}[Latent Gaussian dependence]\label{ass:latent}
For each $n$, units $i=1,\ldots,n$ occupy distinct sites $s_i$ in a fixed countable set $\mathcal D\subset\R^2$ whose points are at mutual distance at least one (after rescaling); write $d_{ik}=\|s_i-s_k\|$. Observations are
\[
O_i=F(\Gamma_i,\varepsilon_i),
\]
where $F$ is measurable and does not depend on $i$ or $n$; $\varepsilon_1,\ldots,\varepsilon_n$ are i.i.d.\ and independent of $\Gamma=(\Gamma_1,\ldots,\Gamma_n)$; and $\Gamma$ is a centred Gaussian array with $\Gamma_i\in\R^r$ and
\[
\Cov(\Gamma_i,\Gamma_k)=c(d_{ik})\,\Sigma_\Gamma ,
\]
for a correlation function $c$ with $0\le c(d)\le C_0e^{-d/\ell}$. The spatial correlation matrices $C_n=\{c(d_{ik})\}$ satisfy $\lambda_{\min}(C_n)\ge\lambda_*>0$ for all $n$.
\end{assumption}

The exposure, covariates and potential outcomes of unit $i$ are all components of $O_i$, so each unit's marginal law is free of the dependence and Sections~\ref{sec:ident}--\ref{sec:effects} apply unit by unit. The model excludes interference: $Y_i(w)$ does not depend on other units' exposures. Because $F$ does not depend on $i$, the assumption also excludes covariates that are deterministic functions of location, such as latitude, and population sizes that vary systematically over space, unless these are themselves modelled as draws from the latent field or the noise. In the application of Section~\ref{sec:application} the assumption is therefore a working model for the dependence of the scores, not a description of how the covariates arose.

\paragraph{The Gaussian-process--Ornstein--Uhlenbeck case}
Let $\Gamma_i=(\Xi_i,u_i)$, where $\Xi_i$ collects the spatially structured parts of the covariates and of a latent policy propensity, and $u_i=(u_{i1},\ldots,u_{iJ})$ is a latent log-incidence trajectory with
\[
\Cov(u_{ij},u_{kl})=\sigma^2c(d_{ik})e^{-\theta|j-l|}.
\]
In time this is an Ornstein--Uhlenbeck process sampled annually, an AR(1) sequence with lag-one correlation $e^{-\theta}$. In space it is an exponential (Mat\'ern-$\tfrac12$) Gaussian process when $c(d)=e^{-d/\ell}$. The OU correlation enters only $\Sigma_\Gamma$, which contains the block $\sigma^2(e^{-\theta|j-l|})_{jl}$. Unit-level nuggets, population sizes and Poisson counting noise belong to $\varepsilon_i$. The requirement $\lambda_*>0$ holds for the exponential kernel on separated sites, because its smallest eigenvalue admits a lower bound that depends only on the separation distance \citep[Chapter~12]{wendland2005scattered}. Several independent separable components with different ranges are handled by Remark~\ref{rem:multirange}.

\begin{lemma}[Exponential $\rho$-mixing]\label{lem:mixing}
Under Assumption~\ref{ass:latent}, for disjoint nonempty index sets $I,J$ with $r=\min_{i\in I,k\in J}d_{ik}$,
\[
\rho\{\sigma(O_I),\sigma(O_J)\}\le\frac{1}{\lambda_*}\sup_{i}\sum_{k:d_{ik}\ge r}c(d_{ik})\le K_0e^{-r/(2\ell)},
\]
and $\alpha\{\sigma(O_I),\sigma(O_J)\}\le\rho\{\sigma(O_I),\sigma(O_J)\}/4$. The constant $K_0$ depends only on $C_0$, $\ell$ and $\lambda_*$, not on $n$, $|I|$ or $|J|$.
\end{lemma}

Here $\rho$ and $\alpha$ are the maximal-correlation and strong-mixing coefficients \citep{bradley2005basic}. The bound is uniform in the cardinalities of $I$ and $J$, which is the form required by central limit theorems for random fields \citep{jenish2009central}. Its proof combines three facts. The maximal correlation of a Gaussian vector equals its maximal linear (canonical) correlation \citep{kolmogorov1960strong}. Independent noise does not increase maximal correlation \citep{csaki1963}. A Schur bound then controls the cross-covariance operator.

\begin{proposition}[Mehler representation of score covariances]\label{prop:mehler}
Under Assumption~\ref{ass:latent}, let $\phi:\mathcal O\to\R^q$ satisfy $\E\phi(O_1)=0$ and $\E\|\phi(O_1)\|^2<\infty$, and put $C_P=\E\phi\phi\tp$. There are positive semidefinite matrices $M_1,M_2,\ldots$ with $\sum_{m\ge1}M_m\preceq C_P$ such that, for all $i\ne k$,
\begin{equation}\label{eq:mehler}
\Cov\{\phi(O_i),\phi(O_k)\}=g\{c(d_{ik})\},\qquad g(c)=\sum_{m\ge1}c^mM_m ,
\end{equation}
the series converging absolutely on $[0,1]$. The matrix function $g$ is nondecreasing and convex on $[0,1]$ in the Loewner order, $g(0)=0$, $g(1)\preceq C_P$, and $g(c)\preceq cC_P$. Consequently
\begin{equation}\label{eq:Csp}
\Csp_n:=n\Var\Bigl(\frac1n\sum_{i=1}^n\phi(O_i)\Bigr)=C_P+\sum_{m\ge1}S_{m,n}M_m\succeq C_P,\qquad
S_{m,n}=\frac1n\sum_{i\ne k}c(d_{ik})^m .
\end{equation}
\end{proposition}
Explicitly, $M_m=\sum_{|\alpha|=m}f_\alpha f_\alpha\tp$, where $f_\alpha$ are the Hermite coefficients of $f(z)=\E\phi\{F(Az,\varepsilon_1)\}$ and $AA\tp=\Sigma_\Gamma$. The first chaos $M_1$ is the part of the score that is linear in the latent field; higher chaoses capture the nonlinearity introduced by counting, logarithms and quantile maps. Since $c\ge0$, every term of \eqref{eq:Csp} beyond $C_P$ is positive semidefinite. Hence independence-based standard errors understate the variance of any functional, and never overstate it. If $\sqrt n\langle h,\widehat\Delta-\Delta\rangle$ is asymptotically normal with variance $h\tp\Csp h$ and $\E\|\phi\|^4<\infty$, the nominal $1-a$ independence interval has asymptotic coverage
\begin{equation}\label{eq:undercover}
2\Phi\bigl(z_{1-a/2}\,r_h\bigr)-1,\qquad r_h^2=\frac{h\tp C_Ph}{h\tp C_Ph+\sum_mS_{m}\,h\tp M_mh}\le1,
\end{equation}
with strict undercoverage whenever some $h\tp M_mh\,S_m>0$; here $S_m=\lim S_{m,n}$ along the sequence.

\begin{remark}\label{rem:multirange}
If $\Gamma$ is the sum of $R$ independent separable components with correlation functions $c_1,\ldots,c_R$, Lemma~\ref{lem:mixing} holds with the slowest decay rate. Proposition~\ref{prop:mehler} holds with $g$ replaced by a power series in $(c_1,\ldots,c_R)$ with positive semidefinite coefficients. The estimators below then require one range per component.
\end{remark}

\subsection{Causal estimating equations under dependence}
We consider a binary exposure $W\in\{0,1\}$, a fixed common sufficient set $S$, an outcome represented in $\R^q$ (for example the level and $q-1$ quantiles of a timing distribution) and parametric nuisance models. For the ATE these are a logistic propensity model in $X_S$ and arm-wise linear outcome regressions. Write $\vartheta=(\beta,\gamma_0,\gamma_1,\mu_0,\mu_1)$ for the propensity, regression and arm-mean parameters, and $g(O,\vartheta)$ for the stacked estimating function. It consists of the logistic score, the two sets of normal equations, and the AIPW mean equations
$\E\{m_v(X_S;\gamma_v)+\I(W=v)(\widetilde Y-m_v)/\pi_v(X_S;\beta)-\mu_v\}=0$. The estimator is $\widehat\Delta=\widehat\mu_1-\widehat\mu_0$. The incremental effect stacks the same nuisance equations with $\E\{\varphi_\delta(O;\beta,\gamma_0,\gamma_1)-\widetilde Y-\tau_\delta\}=0$, using \eqref{eq:incif}. The mediation effects stack the logistic score, a parametric mediator model, outcome regressions on $(X_S,M)$ and the equations $\E(\psi_{ab}-\theta_{ab})=0$ of Theorem~\ref{thm:mediation}. In each case the target $\theta$ is a linear function of the last block of $\vartheta$.
\begin{assumption}[Dependent causal estimating equations]\label{ass:depaipw}
Assumption~\ref{ass:latent} holds, and Assumptions~\ref{ass:unconf} and~\ref{ass:overlap} (for mediation, Assumption~\ref{ass:seqign}) hold for the common marginal law of the units with the set $S$; the relevant potential outcomes have finite second moments. The pseudo-true $\vartheta_0$ is the unique zero of $\E g(O_1,\vartheta)$, lies in the interior of a compact set $\Theta$, and its target component equals the causal estimand $\theta$. This holds for the ATE if at least one nuisance model is correct, for the incremental effect if the propensity model is correct, and for mediation effects if at least two of the three models are correct (Theorems~\ref{thm:causalid} and~\ref{thm:mediation}). On $\Theta$, $g$ is continuously differentiable in $\vartheta$, and for some $\kappa>0$, $\E\sup_\Theta\|g(O_1,\vartheta)\|^{4+\kappa}$ and $\E\sup_\Theta\|\partial_\vartheta g(O_1,\vartheta)\|^{2}$ are finite. The Jacobian $G=\E\partial_\vartheta g(O_1,\vartheta_0)$ is nonsingular, and $\liminf_n\lambda_{\min}(\Csp_n)>0$ for the score $\phi$ defined below.
\end{assumption}

\begin{theorem}[Central limit theorem under latent Gaussian dependence]\label{thm:depclt}
Under Assumption~\ref{ass:depaipw}, let $\phi(O)$ be the target component of $-G^{-1}g(O,\vartheta_0)$, the stacked influence function; for the ATE it is the $\mu_1-\mu_0$ component. Then
\[
\widehat\theta-\theta=\frac1n\sum_{i=1}^n\phi(O_i)+o_P(n^{-1/2}),\qquad
(\Csp_n)^{-1/2}\sqrt n(\widehat\theta-\theta)\Rightarrow N(0,I_q),
\]
with $\Csp_n$ given by \eqref{eq:Csp} for this $\phi$.
\end{theorem}
Thus the ATE, incremental and mediation estimators share one limit theory, and the covariance estimators below apply to each. The stacked influence function includes the estimation effects of all nuisance models. The plug-in score $\widehat\psi_1-\widehat\psi_0-\widehat\Delta$ omits them, and is adequate only when both models are correct. Cross-fitting is not used here because sample splitting does not produce independent folds under dependence.

\subsection{Estimating the long-run covariance}\label{sec:covest}
Write $\widehat\phi_i=\phi(O_i;\widehat\vartheta)$, $\widetilde\phi_i=\widehat\phi_i-\overline{\widehat\phi}$, $\widehat C_P=n^{-1}\sum_i\widetilde\phi_i\widetilde\phi_i\tp$, and $\widehat c_{ik}=e^{-d_{ik}/\widehat\ell}$, where $\widehat\ell$ estimates the range $\ell$ of the latent correlation $c(d)=e^{-d/\ell}$ in Assumption~\ref{ass:latent}. Because that correlation is shared by all latent components, a correctly specified latent Gaussian model for the outcome identifies it; we use the Poisson GP--OU model of Section~\ref{sec:application} for counts. A Gaussian restricted-maximum-likelihood fit to a transformed outcome is only a working estimator: by Proposition~\ref{prop:mehler} the covariance of a nonlinear transform is a series in $c$ rather than an exponential. Part~(d) of the next theorem shows that overstating the range is harmless for the chord estimator.

\paragraph{Mehler estimator} Truncate \eqref{eq:mehler} at order $m_0$. For pairs $\mathcal P_n=\{(i,k):i<k,\ d_{ik}\le h\}$, regress the products $\tfrac12(\widetilde\phi_i\widetilde\phi_k\tp+\widetilde\phi_k\widetilde\phi_i\tp)+\widehat C_P/n$ on the regressors
\[
a_{ik}^{(m)}=\widehat c_{ik}^{\,m}-\{\widehat r_i^{(m)}+\widehat r_k^{(m)}-\widehat S_m\}/n,\qquad
\widehat r_i^{(m)}=\sum_{j\ne i}\widehat c_{ij}^{\,m},\quad \widehat S_m=\frac1n\sum_i\widehat r_i^{(m)},
\]
for $m=1,\ldots,m_0$, minimizing the sum of squared Frobenius residuals subject to $\widehat M_m\succeq0$ for every $m$ (a convex problem, solved by projected gradient), and set
\begin{equation}\label{eq:Vmehler}
\widehat V_{\rm M}=\Bigl(\widehat C_P+\sum_{m=1}^{m_0}\widehat S_m\widehat M_m\Bigr)\Big/\Bigl(1-\frac1n\Bigr).
\end{equation}
The corrections involving $1/n$ are exact for products of mean-centred scores. They remove a downward bias of order $\Csp_n/n$, which matters when dependence makes the effective sample size small.

\paragraph{Convex-chord estimator} Group pairs with $d_{ik}\le h$ into distance bins $B_j=\{(i,k):d_{j-1}<d_{ik}\le d_j\}$, $j=1,\ldots,K$, with edges $0=d_0<\cdots<d_K=h$ fixed in advance. Let $\bar c_j$ be the bin mean of $\widehat c_{ik}$, and let $\widehat g_j$ be the bin mean of the centred products plus $\widehat V_{\rm ch}/n$, projected onto the positive semidefinite cone. The term $\widehat V_{\rm ch}/n$ is an approximate finite-sample centering adjustment: for mean-zero scores, $\E\{(\phi_i-\bar\phi)(\phi_k-\bar\phi)\tp\}=C_{ik}-(R_i+R_k\tp-V_n)/n$ with $C_{ik}=\E\phi_i\phi_k\tp$, $R_i=\sum_jC_{ij}$ and $V_n=n^{-1}\sum_{i,j}C_{ij}$, so adding $V_n/n$ is exact only when all row sums $R_i$ equal $V_n$, unlike the pair-specific correction of the Mehler estimator. It is asymptotically negligible; because $\widehat V_{\rm ch}$ enters its own knots, it is computed as a fixed point. Let $\widehat{\rm ch}$ be the piecewise-linear matrix function on $[0,1]$ through $(0,0)$, the points $(\bar c_j,\widehat g_j)$ and $(1,\widehat C_P)$. Set
\begin{equation}\label{eq:Vchord}
\widehat V_{\rm ch}=\Bigl\{\widehat C_P+\frac2n\sum_{i<k}\widehat{\rm ch}(\widehat c_{ik})\Bigr\}\Big/\Bigl(1-\frac1n\Bigr),
\end{equation}
projected onto the positive semidefinite cone (which, since the knots and $\widehat C_P$ are positive semidefinite and the chord weights nonnegative, changes nothing in exact arithmetic).
Because $g$ is convex with $g(0)=0$ and $g(1)\preceq C_P$, and bin means of a convex function lie above its graph by Jensen's inequality, the chord dominates $g$ whatever the chaos order. This makes $\widehat V_{\rm ch}$ asymptotically conservative. Pairs beyond $h$ enter both estimators through the parametric correlation, unlike kernel estimators that truncate them.

\begin{theorem}[Covariance estimation]\label{thm:covest}
Suppose Assumption~\ref{ass:depaipw} holds with $c(d)=e^{-d/\ell}$, $\widehat\ell\to\ell$ in probability, and $\phi(o;\vartheta)$ Lipschitz in $\vartheta$ near $\vartheta_0$ with a square-integrable constant. Fix $h\ge1$ and let every pair set used have cardinality at least a constant multiple of $n$.
\begin{enumerate}[label=(\alph*),leftmargin=1.6em,itemsep=2pt]
\item If $M_m=0$ for $m>m_0$ and the Gram matrices $|\mathcal P_n|^{-1}\sum_{\mathcal P_n}a_{ik}a_{ik}\tp$, with $a_{ik}=(c_{ik},\ldots,c_{ik}^{m_0})$, have smallest eigenvalues bounded away from zero, then $\widehat V_{\rm M}-\Csp_n\to0$ in probability.
\item For fixed distance bins whose mean correlations remain separated, for every $\lambda\in\R^q$ and $\epsilon>0$, $P(\lambda\tp\widehat V_{\rm ch}\lambda\ge\lambda\tp\Csp_n\lambda-\epsilon)\to1$. Moreover $\widehat V_{\rm ch}-\Csp_n\preceq E_n+o_P(1)$, where
\[
E_n=\frac2n\sum_{i<k}\{{\rm ch}_n(c_{ik})-g(c_{ik})\}\succeq0
\]
is the deterministic chord error of the population bin means.
\item Consequently, Wald intervals and $L^2$ tests based on $\widehat V_{\rm M}$ have asymptotically exact level under (a), and those based on $\widehat V_{\rm ch}$ have asymptotic coverage at least nominal under (b).
\item The conclusions of (b) and (c) for $\widehat V_{\rm ch}$ remain true if $\widehat\ell\to\ell'$ in probability for some $\ell'\ge\ell$, with $E_n$ computed for the correlations $e^{-d_{ik}/\ell'}$.
\end{enumerate}
\end{theorem}
The chord error $E_n$ is small when bins are narrow where the pairs are dense. For the exponential correlation on planar sites, $S_{m,n}$ is approximately $m^{-2}S_{1,n}$ when the range is long relative to the spacing, and decays geometrically in $m$ when it is short; higher chaoses therefore contribute little unless the score is strongly nonlinear in the latent field. Kernel heteroskedasticity-and-autocorrelation-consistent (HAC) estimators \citep{kelejian2007hac} and block estimators remain consistent under the mixing of Lemma~\ref{lem:mixing} with suitable bandwidths. With few units within the dependence range they are strongly biased downwards, as the experiments in Section~\ref{sec:numerical} show. Part~(d) suggests using a deliberately generous range with the chord estimator when the range is uncertain; the price is a wider interval.

\paragraph{What the theorem covers} Theorem~\ref{thm:covest} concerns the estimators as defined above: fixed distance bins and, for the Mehler estimator, a known finite chaos order. Our default implementation of the chord estimator instead uses quantile bins of $\widehat c_{ik}$ within the cutoff $\widehat c_{ik}\ge0.02$, which depends on $\widehat\ell$, merges sparse bins and iterates the centering correction twice (Supplement Section~S1). This adaptive version is not covered by the theorem, and neither is the implemented Mehler pair set, whose cutoff $h=\max(4\widehat\ell,3\bar d_{\rm nn})$ also depends on $\widehat\ell$. We therefore also report the fixed-bin estimator, with bins chosen before seeing the data; in the stress test, the mediation simulation and the applications the two versions give similar standard errors and coverage (Supplement Tables~S4, S7 and~S10). For scores that are nonlinear in the latent field, such as those of counts, logarithms and quantiles, the chaos order is not finite, and the Mehler estimator with $m_0=2$ is an approximation whose omitted terms are not bounded; Supplement Tables~S4, S7 and~S10 compare $m_0=1,2,3$. The chord estimator does not depend on the chaos order. All guarantees are asymptotic. With few effectively independent units the variance estimators are themselves variable, and even the chord interval can fall below nominal coverage (Section~S3).

\subsection{Collections under dependence: block fractional sampling}
The collection of Section~\ref{sec:collection} resamples units; under dependence, entire spatial blocks are resampled instead.
\begin{assumption}[Blocks]\label{ass:blocks}
Assumption~\ref{ass:latent} holds, and the sites are partitioned into $G_n$ blocks of $m_n$ units each, with $m_n\to\infty$ and $m_n/n\to0$. Each block is contained in a square of side at most $C_Bm_n^{1/2}$, and for every $t\ge0$ at most $C_B(1+t)m_n^{1/2}$ of its units lie within distance $t$ of a unit outside the block. Square tiles of a lattice satisfy both conditions. The tensor satisfies \eqref{eq:Tlinear} with an influence function $\psi_T$ of the latent model. For a random union $U$ of $G_n/2$ blocks drawn without replacement, independent of the data, $\widehat{\mathbf T}_U-\mathbf T=|U|^{-1}\sum_{i\in U}\psi_T(O_i)+R_U$ with $\sqrt n\|R_U\|\to0$ in probability. The block sums $V_g=\sum_{i\in g}\psi_T(O_i)$ satisfy $\sup_g\E\|V_g\|^4\le C m_n^2$, and $\Csp_{T,n}=n\Var(\sum_i\psi_T(O_i)/n)\to\Csp_T$.
\end{assumption}
The linearization of $\widehat{\mathbf T}$ under dependence is assumed here, not proved; Theorem~\ref{thm:rate} establishes it only for independent data. The fourth-moment block bound can be derived from the Rosenthal-type inequality of \citet{bryc1993moment} for families whose maximal correlation between disjoint index sets is bounded below one. Under Assumption~\ref{ass:latent} that correlation is at most $(\kappa_n-1)/(\kappa_n+1)$, where $\kappa_n$ is the bounded condition number of $C_n$, by the argument of Lemma~\ref{lem:mixing} and Wielandt's inequality; we keep the bound as an assumption.

\begin{theorem}[Block fractional sampling]\label{thm:blockhalf}
Under Assumption~\ref{ass:blocks}, let $Z_U=\sqrt n\,\mathcal N(\widehat{\mathbf T}_U-\widehat{\mathbf T}_n)$, the case $|U|=n/2$ of \eqref{eq:ZI}. Its conditional distribution given the data converges in probability to that of $\mathcal N(Z)$, $Z\sim N(0,\Csp_T)$. If this law is continuous and strictly increasing at its $(1-\gamma)$ quantile, the conclusions of Theorem~\ref{thm:subsample} hold for the collection calibrated by $Z_U$.
\end{theorem}
Unit-level fractional sampling is the special case of single-unit blocks under independence; there the remainder condition follows from Theorem~\ref{thm:rate}.

\section{Numerical studies}\label{sec:numerical}
All results in this section come from the author's code, which is available on request; seeds and replication counts are fixed in the scripts, and Section~S2 of the Supplementary Material gives every design in full. A finite simulation cannot verify an asymptotic theorem or global optimization; Monte Carlo intervals are reported where coverage is the target. A rate study of the projector estimator under Gaussian, $t_{2.5}$ and contaminated errors (Table~S1) gives log--log slopes between $-0.52$ and $-0.57$ for $n$ from 100 to 800, consistent with Theorem~\ref{thm:rate}.

\subsection{Calibration of the outer collection}\label{sec:sim-collection}
We compare conventional subsampling with $b=n^{0.7}$ and scaling $\sqrt b$ with fractional sampling at $b=n/2$ and scaling $\sqrt n$, both computed on the same full-sample fit with 99 resamples and nominal simultaneous coverage $0.90$. In the strong-signal design, $X\mid W\sim N\{W(.8,.8,0)\tp,I_3\}$ and the potential response is the uniform distribution on $[-R(w),R(w)]$ with $R(w)=\exp(.3X_1+.12e)+.15w$. Its mean is always zero, so confounding acts only on shape, and the sufficient collection consists of the four supersets of $\{1\}$. The weak-signal design reduces the confounder's mean shift to $0.2$ and raises the radius noise to $0.8$. The spatial design places the weak-signal design on a $20\times20$ lattice with latent exponential range two.
\begin{table}[t]\centering\small
\caption{Simultaneous outer collection, nominal coverage $0.90$, 200 replications per row and 99 resamples per dataset. Coverage requires every sufficient set in both arms; the interval is the Wilson 95\% Monte Carlo interval. The true armwise collection has four of eight sets. The spatial design uses a $20\times20$ lattice with latent exponential range two and blocks of $5\times5$ units. The rows with the finite-population factor reuse the subsampling quantiles multiplied by $\{n/(n-b)\}^{1/2}$, which isolates the effect of the scaling from that of the subset size; exactness and size were not recorded for them.}\label{tab:collection}
\begin{tabular}{llllrrr}\toprule
Design & $n$ & Calibration & Coverage & MC interval & Exact & Size\\\midrule
Strong signal & 200 & Subsampling, $b=n^{0.7}$ & 1.000 & [0.981, 1.000] & 0.995 & 4.00\\
Strong signal & 200 & Subsampling, $b=n^{0.7}$, finite-pop. factor & 1.000 & [0.981, 1.000] & -- & --\\
Strong signal & 200 & Fractional, $b=n/2$ & 1.000 & [0.981, 1.000] & 0.995 & 4.01\\
Strong signal & 400 & Subsampling, $b=n^{0.7}$ & 1.000 & [0.981, 1.000] & 1.000 & 4.00\\
Strong signal & 400 & Subsampling, $b=n^{0.7}$, finite-pop. factor & 1.000 & [0.981, 1.000] & -- & --\\
Strong signal & 400 & Fractional, $b=n/2$ & 1.000 & [0.981, 1.000] & 1.000 & 4.00\\
Weak signal & 200 & Subsampling, $b=n^{0.7}$ & 0.685 & [0.618, 0.745] & 0.025 & 6.79\\
Weak signal & 200 & Subsampling, $b=n^{0.7}$, finite-pop. factor & 0.800 & [0.739, 0.850] & -- & --\\
Weak signal & 200 & Fractional, $b=n/2$ & 0.995 & [0.972, 0.999] & 0.000 & 7.94\\
Weak signal & 400 & Subsampling, $b=n^{0.7}$ & 0.780 & [0.718, 0.832] & 0.055 & 6.82\\
Weak signal & 400 & Subsampling, $b=n^{0.7}$, finite-pop. factor & 0.855 & [0.800, 0.897] & -- & --\\
Weak signal & 400 & Fractional, $b=n/2$ & 0.980 & [0.950, 0.992] & 0.005 & 7.82\\
Weak, spatial & 400 & Fractional, units & 0.950 & [0.910, 0.973] & 0.030 & 7.51\\
Weak, spatial & 400 & Fractional, blocks & 0.980 & [0.950, 0.992] & 0.005 & 7.79\\
\bottomrule\end{tabular}\end{table}

With a strong signal both calibrations cover and recover the exact collection in at least 99.5\% of datasets, so fractional sampling costs no power (Table~\ref{tab:collection}). With a weak signal, subsampling undercovers ($0.685$ and $0.780$ at $n=200,400$), whereas fractional sampling covers ($0.995$ and $0.980$). Applying the finite-population factor $\{n/(n-b)\}^{1/2}$ to the same subsampling quantiles raises coverage to $0.800$ and $0.855$, so about a third of the improvement comes from the scaling and the rest from the larger subsets, whose projector errors do not saturate. The fractional collections retain 7.8--7.9 of the eight candidate sets, four of which are insufficient. They therefore provide conservative protection against excluding a sufficient set, not evidence that the sufficient sets have been recovered; the data cannot exclude the insufficient sets at these sample sizes. In the spatial design, unit-level fractional sampling covers $0.950$ and block fractional sampling of $5\times5$ blocks, as prescribed by Theorem~\ref{thm:blockhalf}, covers $0.980$.

\subsection{Functional effects under latent GP--OU dependence}\label{sec:sim-dependence}
Units lie on a square lattice ($n=196,400,900$) or at the 49 contiguous U.S.\ state centroids. A policy field, four covariate fields and a six-year latent log-incidence trajectory are independent Gaussian fields with spatial correlation $e^{-d/\ell}$; the trajectory has OU correlation $e^{-0.5|j-l|}$ across years. Exposure is $W=\I(V>0)$ for the policy field $V$, so it is spatially clustered, and $X\mid W$ is Gaussian with mean shift $(.7,.6,0,.8)W$. Thus $X_1,X_2$ are confounders, $X_3$ affects only the outcome and $X_4$ only the exposure. Annual Poisson counts have a level depending on $(W,X_1,X_3)$ and the latent field, and a Beta timing law whose mean depends on $(W,X_1,X_3)$ and whose concentration depends on $(W,X_2)$. The outcome is the level, the mean timing and ten timing quantiles. AIPW adjusts for $\{X_1,X_2,X_3\}$ with a logistic propensity, correctly specified here, and linear outcome regressions, which are misspecified. The range $\widehat\ell$ is the REML estimate for the level regressed on $(1,W,X)$.

\begin{table}[t]\centering\footnotesize
\caption{Coverage of nominal 95\% intervals for the level effect under the latent GP--OU model (400 replications per lattice row, 800 per state row). $\widehat\ell$ is the median REML range; the last column is the coverage of the $L^2$ confidence set for the timing quantile map using the chord estimator. HAC uses a Bartlett kernel with bandwidth $2n^{1/4}$ grid units (1200 km for states); blocks are squares of about $1.2n^{1/4}$ units (eight $k$-means blocks for states). $\ell=0$ is independence.}\label{tab:dependence}
\setlength{\tabcolsep}{4pt}\begin{tabular}{lrrrrrrrrrrr}\toprule
Geometry & $n$ & $\ell$ & $\widehat\ell$ & $|$bias$|$/sd & i.i.d. & HAC & Block & Mehler & Chord & Chord, true $\ell$ & $L^2$\\\midrule
Lattice & 196 & 0 & 0.173 & 0.01 & 0.965 & 0.895 & 0.935 & 0.970 & 0.968 & -- & 0.973\\
Lattice & 196 & 1 & 1.06 & 0.01 & 0.863 & 0.833 & 0.887 & 0.948 & 0.988 & 0.990 & 0.990\\
Lattice & 196 & 2 & 2 & 0.03 & 0.772 & 0.810 & 0.835 & 0.943 & 0.990 & 0.988 & 0.990\\
Lattice & 400 & 0 & 0.05 & 0.08 & 0.945 & 0.910 & 0.922 & 0.948 & 0.950 & -- & 0.945\\
Lattice & 400 & 1 & 1.04 & 0.06 & 0.890 & 0.897 & 0.912 & 0.955 & 0.983 & 0.988 & 0.985\\
Lattice & 400 & 2 & 2 & 0.07 & 0.693 & 0.825 & 0.850 & 0.935 & 0.973 & 0.973 & 0.973\\
Lattice & 900 & 0 & 0.143 & 0.03 & 0.945 & 0.905 & 0.925 & 0.953 & 0.958 & -- & 0.958\\
Lattice & 900 & 1 & 1.01 & 0.00 & 0.900 & 0.927 & 0.932 & 0.978 & 0.995 & 0.985 & 0.995\\
Lattice & 900 & 2 & 1.95 & 0.11 & 0.690 & 0.855 & 0.870 & 0.940 & 0.963 & 0.958 & 0.963\\
States & 49 & 0 km & 49 & 0.07 & 0.890 & 0.845 & 0.864 & 0.902 & 0.925 & -- & 0.925\\
States & 49 & 260 km & 272 & 0.02 & 0.870 & 0.855 & 0.858 & 0.924 & 0.976 & 0.986 & 0.979\\
States & 49 & 520 km & 534 & 0.01 & 0.810 & 0.811 & 0.816 & 0.925 & 0.986 & 0.990 & 0.990\\
\bottomrule\end{tabular}\end{table}

\begin{figure}[t]
\centering
\includegraphics[width=.92\textwidth]{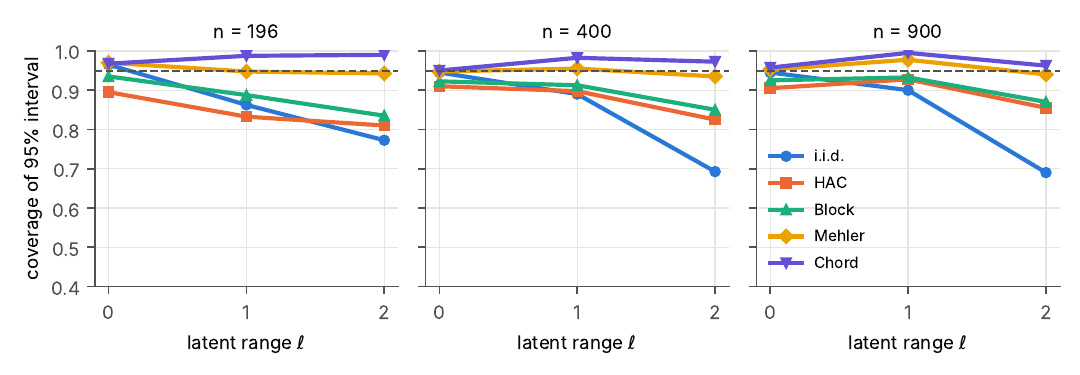}
\caption{Coverage of nominal 95\% intervals for the level effect on lattices, by covariance estimator and latent range (400 replications per point; Monte Carlo standard error about 0.011 near 0.95). Plotted directly from the stored results by the analysis code (see the generative-AI declaration).}\label{fig:dependence}
\end{figure}

Table~\ref{tab:dependence} and Figure~\ref{fig:dependence} show the consequences of Proposition~\ref{prop:mehler}. The AIPW estimator is essentially unbiased, with absolute bias at most 0.11 Monte Carlo standard deviations. Under independence every interval has close to nominal coverage. As the range grows, independence intervals undercover, falling to $0.69$ at $\ell=2$; their standard errors are then about half the Monte Carlo standard deviation, as \eqref{eq:undercover} predicts. Kernel HAC and block intervals cover only $0.81$--$0.93$, because with few units per correlation range they truncate or block away much of the positive covariance. The Mehler standard error is on average within 11\% of the Monte Carlo standard deviation in every lattice cell, and its intervals cover $0.94$--$0.98$. Its shortfall at $\ell=2$ reflects the variability of the variance estimate when only a few dozen effectively independent units are available. The chord intervals cover $0.95$--$0.995$, as Theorem~\ref{thm:covest}(b) leads one to expect, with standard errors 12--34\% above the Monte Carlo standard deviation. Using the estimated rather than the true range costs nothing visible, the $L^2$ confidence set for the timing quantile map behaves like the chord interval, and results for the mean timing are similar (Table~S2).

With the 49 state centroids and ranges of 0, 260 and 520 km, the last two bracketing the fitted range of Section~\ref{sec:application}, independence, HAC (1200 km bandwidth) and block intervals (eight $k$-means blocks) cover only $0.81$--$0.89$, even under independence, where HAC and blocks lose degrees of freedom. The Mehler intervals cover $0.90$--$0.93$ and the chord intervals $0.93$--$0.99$. With fewer than 50 units and nonzero range, only the chord estimator attains nominal coverage; this is the setting of the state analysis. These favourable results should not be read as finite-sample guarantees. In a stress test with Gaussian-field scores and 40--50 effectively independent units (Section~S3), the chord intervals cover 0.89--0.91 with the true range, 0.90--0.93 with fixed distance bins, and 0.91--0.95 with twice the range.

\subsection{Difference-in-differences with dependent errors}\label{sec:sim-panel}
To check the inference of Section~\ref{sec:application}, we generate 300 panels from the GP--OU model fitted to the 44 jurisdictions for 2008--2019, with the observed populations, the observed 2014 cohort, the fitted state and year effects, a zero expansion effect, and latent errors with the fitted covariance (range 172 km, lag-one correlation 0.94). Both estimators in Table~\ref{tab:panel} are unbiased. The two-way Poisson estimator has Monte Carlo standard deviation 0.29, consistent with its weighting of jurisdictions by their counts, which lets a few large states with persistent trajectories dominate. Its model-based, quasi-Poisson and state-cluster intervals cover 0.03, 0.31 and 0.78 of the time. The GP--OU mixed model, which re-estimates all hyper-parameters, has standard deviation 0.087, a 3.3-fold reduction; its mean reported standard error equals that standard deviation and its intervals cover 0.94. Cluster-robust inference with 44 clusters therefore does not rescue the two-way analysis when incidence is as persistent as the fitted OU process implies. Because these panels come from the mixed model itself, the comparison favours it by construction. In a misspecified design with state-specific linear trends and independent noise (Supplement Table~S5), the two-way state-cluster intervals cover 0.84, and the mixed model is unbiased with intervals covering 0.94.
\begin{table}[t]\centering\small
\caption{Difference-in-differences with GP--OU latent errors calibrated to the fitted state panel (44 jurisdictions, 2008--2019, true effect zero, 300 replications): bias, Monte Carlo standard deviation, mean reported standard error and coverage of nominal 95\% intervals.}\label{tab:panel}
\begin{tabular}{lrrrr}\toprule
Estimator & Bias & SD & Mean SE & Coverage\\\midrule
Two-way Poisson, model SE & 0.006 & 0.286 & 0.008 & 0.033\\
Two-way Poisson, quasi-Poisson SE & 0.006 & 0.286 & 0.056 & 0.313\\
Two-way Poisson, state-cluster SE & 0.006 & 0.286 & 0.182 & 0.783\\
GP--OU Poisson mixed model & -0.002 & 0.087 & 0.087 & 0.940\\
\bottomrule\end{tabular}\end{table}

\subsection{Mediation and incremental effects under dependence}\label{sec:sim-mediation}
On a $20\times20$ lattice, independent Gaussian fields with range $\ell\in\{0,1,2\}$ drive a spatially clustered exposure whose propensity given $X$ is exactly logistic, a Gaussian linear mediator and two outcomes (Section~S2). The outcome regression for the first outcome is misspecified through a covariate--mediator interaction. The mediation estimator is then consistent by Theorem~\ref{thm:mediation}(b), because the propensity and mediator models are correct, and the incremental estimator because the propensity model is correct. The range is the REML estimate for the mediator, a correctly specified Gaussian latent model, with medians 1.00 and 1.90 for $\ell=1,2$. Table~\ref{tab:medsim} reports the first outcome, with 400 replications per range. All estimators are essentially unbiased. Under independence, the independence, Mehler and chord intervals cover 0.92--0.98 and HAC intervals 0.89--0.93; the natural indirect effect is slightly below nominal because the mediator density ratio makes its score heavy tailed at $n=400$. As the range grows, independence intervals fall to 0.81--0.85 at $\ell=2$, and HAC and block intervals to 0.85--0.88. For $\ell>0$ the Mehler intervals cover 0.92--0.96 and the chord intervals 0.97--0.98, and with twice the estimated range the chord intervals cover at least 0.985, in line with Theorem~\ref{thm:covest}(d). The chord estimator with fixed distance bins, which Theorem~\ref{thm:covest} covers exactly, covers 0.97--0.99 in every cell, and the Mehler estimator gives almost the same coverage with truncation order $m_0=1,2,3$ (Supplement Table~S7). The second outcome, which is nonlinear in the latent field, behaves similarly (Supplement Table~S6). The covariance theory therefore carries over to the multiply robust mediation and incremental scores without change.
\begin{table}[t]\centering\footnotesize
\caption{Natural direct and indirect effects and the incremental effect ($\delta=2$) under latent spatial dependence on a $20\times20$ lattice (400 replications per range): bias, Monte Carlo standard deviation and coverage of nominal 95\% intervals by covariance estimator. The outcome regression is misspecified; propensity and mediator models are correct. Chord, fixed uses distance bins fixed in advance, as in Theorem~\ref{thm:covest}; Chord, $2\widehat\ell$ uses twice the estimated range.}\label{tab:medsim}
\setlength{\tabcolsep}{4pt}
\begin{tabular}{llrrrrrrrrr}\toprule
$\ell$ & Effect & Bias & SD & i.i.d. & HAC & Block & Mehler & Chord & Chord, fixed & Chord, $2\widehat\ell$\\\midrule
0 & Natural direct & -0.009 & 0.134 & 0.943 & 0.912 & 0.930 & 0.958 & 0.963 & 0.978 & 0.968\\
0 & Natural indirect & 0.012 & 0.114 & 0.917 & 0.890 & 0.912 & 0.930 & 0.932 & 0.975 & 0.945\\
0 & Incremental, $\delta=2$ & -0.001 & 0.018 & 0.965 & 0.930 & 0.945 & 0.973 & 0.975 & 0.990 & 0.983\\
\midrule
1 & Natural direct & 0.004 & 0.153 & 0.927 & 0.897 & 0.915 & 0.955 & 0.980 & 0.990 & 0.988\\
1 & Natural indirect & 0.011 & 0.121 & 0.920 & 0.895 & 0.902 & 0.945 & 0.970 & 0.983 & 0.985\\
1 & Incremental, $\delta=2$ & 0.000 & 0.019 & 0.922 & 0.907 & 0.930 & 0.950 & 0.980 & 0.988 & 0.990\\
\midrule
2 & Natural direct & 0.004 & 0.199 & 0.848 & 0.865 & 0.877 & 0.945 & 0.983 & 0.980 & 0.995\\
2 & Natural indirect & 0.005 & 0.160 & 0.807 & 0.845 & 0.875 & 0.917 & 0.965 & 0.973 & 0.988\\
2 & Incremental, $\delta=2$ & -0.005 & 0.025 & 0.810 & 0.850 & 0.880 & 0.943 & 0.978 & 0.985 & 0.985\\
\bottomrule\end{tabular}\end{table}

In an independent design (Section~S3), discarding shape information is costly: selection on a mean summary has integrated error 29 versus 0.20 $\times10^{-3}$ for the selected set. Omitting exposure-only covariates reduces the error of AIPW, and correctly specified g-computation remains best, so no blanket superiority claim is made.

\section{Syphilis surveillance and Medicaid expansion}\label{sec:application}
We use the public CDC AtlasPlus export of primary and secondary (P\&S) syphilis counts and rates \citep{cdcatlasplus} and the Medicaid expansion dates of the replication data of \citet{baker2025difference}; Section~S4 of the Supplementary Material documents the files and cohorts. The question is whether the 2014 Medicaid expansion cohort differs from non-expanding states in the amount or the timing of P\&S syphilis. The comparison is ecological and observational. The analyses illustrate the estimands and the consequences of dependence for inference; they do not establish that the covariates remove policy confounding.

\paragraph{Targets} For nonnegative annual rates $r_1,\ldots,r_J$ with $\Lambda=\sum_jr_j>0$, the level is $L=\log\Lambda$ and the timing law places mass $p_j=r_j/\Lambda$ uniformly within year $j$. Its quantile function $Q$ and mean $\sum_jp_j(j-1/2)$ are unchanged when all rates are multiplied by $c>0$, while $L$ shifts by $\log c$. Timing contrasts are therefore invariant to exposure-dependent, unit-specific multiplicative ascertainment, provided the multiplier is constant across the window. No pseudo-rate offset is added, because an offset destroys this invariance. Under a log-additive representation $r_{ij}(w)=\exp\{a_i(w)+b_j(w)\}$, equal timing laws are equivalent to $b_j(1)-b_j(0)$ being constant in $j$, which links the timing contrast to event-study dynamics without making it a difference-in-differences identification argument.

\subsection{A Poisson GP--OU latent model}
For counts $y_{it}$ with population $N_{it}$ in jurisdiction $i$ and year $t$,
\begin{equation}\label{eq:gpou}
y_{it}\sim\mathrm{Poisson}\{N_{it}\exp(\eta_{it})\},\qquad \eta_{it}=f_{it}\tp\beta+u_{it},\qquad
u\sim N\bigl\{0,\ \sigma^2C_s(\ell)\otimes R(\theta)+\tau^2I\bigr\},
\end{equation}
where $C_s(\ell)=\{e^{-d_{ik}/\ell}\}$ uses great-circle distances between state centroids and $R(\theta)=\{e^{-\theta|t-t'|}\}$ is the Ornstein--Uhlenbeck correlation. The fixed effects $\beta$ have vague Gaussian priors. The hyper-parameters are estimated by maximizing the Laplace approximation to the marginal likelihood \citep{tierney1986accurate,rue2009approximate}, as in spatio-temporal disease mapping \citep{knorrheld2000bayesian}. Table~\ref{tab:gpou} reports two fits.

\begin{table}[t]\centering\small
\caption{Poisson GP--OU latent models for state P\&S syphilis counts. Descriptive model: 51 jurisdictions, 2000--2019, year effects. Difference-in-differences model: 44 jurisdictions, 2008--2019, state and year effects, effect of the 2014 expansion cohort after 2013 on the log rate. Standard errors in parentheses.}\label{tab:gpou}
\begin{tabular}{lrr}\toprule
 & Descriptive & DiD\\\midrule
Latent standard deviation $\sigma$ & 0.99 & 0.65\\
Spatial range $\ell$ (km) & 260 & 172\\
OU lag-one correlation $e^{-\theta}$ & 0.957 & 0.939\\
Nugget standard deviation $\tau$ & 0.059 & 0.049\\
\midrule
$\Delta$ log marginal likelihood vs.\ independent & 929.3 & --\\
\quad vs.\ time only (OU) & 20.0 & --\\
\quad vs.\ space only (GP) & 785.0 & --\\
\midrule
Expansion effect, two-way Poisson (cluster SE) & -- & 0.166 (0.106)\\
\quad model-based / quasi-Poisson SE & -- & (0.009) / (0.039)\\
Expansion effect, GP--OU mixed model & -- & -0.033 (0.088)\\
\bottomrule\end{tabular}\end{table}

The descriptive fit uses all 51 jurisdictions for 2000--2019 with year effects. Deviations from the national trend are large ($\sigma\approx1$) and extremely persistent: the lag-one OU correlation is $0.96$, so state departures last for decades. They are also spatially correlated, with a range of about 260 km, comparable to the distance between neighbouring state centroids. Removing the temporal component lowers the log marginal likelihood by 785 units, and removing the spatial component lowers it by 20 units. The temporal persistence is the dominant feature, but both components are clearly supported.

\subsection{Difference-in-differences with dependent errors}
The second fit in Table~\ref{tab:gpou} restricts to the 44 jurisdictions that expanded during 2014 (27) or not through 2019 (17), for 2008--2019. It adds state and year effects and an expansion indicator for the 2014 cohort after 2013. A Poisson two-way fixed-effects regression gives an effect of $0.166$ on the log rate. Its model-based standard error ($0.009$) ignores overdispersion; the Pearson dispersion is 19. The quasi-Poisson standard error ($0.039$) ignores dependence over time. Only the state-cluster standard error ($0.106$) accounts for the persistence, and then the estimate is not significant. The GP--OU mixed model estimates $-0.033$ (SE $0.088$). Both coefficients target the effect on the 2014 cohort under parallel trends in log rates, an identifying assumption different from Assumption~\ref{ass:unconf}. One reason the two point estimates differ is weighting: the two-way Poisson fit weights jurisdictions by their counts, so the largest states dominate, whereas the mixed model attributes persistent state-specific deviations to the latent field. Such sensitivity of fixed effects to correlated random effects is a known phenomenon \citep{hodges2010adding,paciorek2010importance}.

\begin{figure}[t]
\centering
\includegraphics[width=.75\textwidth]{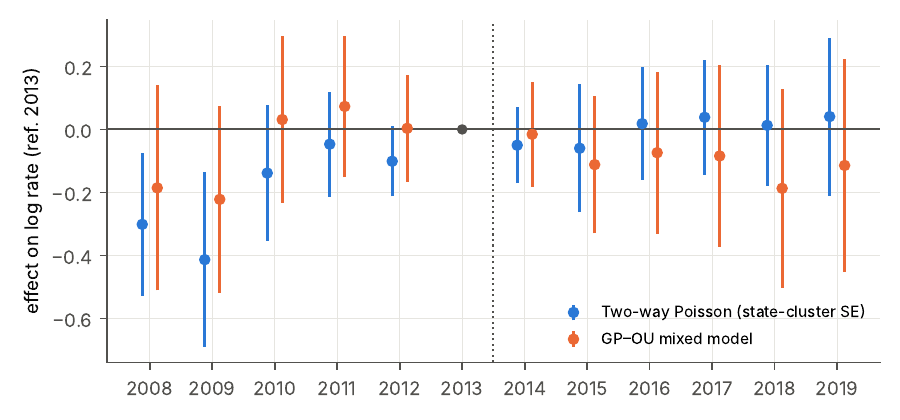}
\caption{Event-study coefficients for the 2014 expansion cohort relative to 2013, with 95\% intervals: Poisson two-way fixed effects with state-cluster standard errors, and the GP--OU Poisson mixed model. Plotted directly from the stored results by the analysis code (see the generative-AI declaration).}\label{fig:event}
\end{figure}

Figure~\ref{fig:event} shows the event-study version. With two-way fixed effects, the 2008 and 2009 leads are $-0.30$ and $-0.41$, with cluster-based intervals excluding zero, which signals non-parallel pre-trends. The mixed model's leads are smaller ($-0.19$, $-0.22$) and its intervals include zero, consistent with the OU component absorbing slowly varying state-specific departures. Nonsignificant leads under a model that absorbs such departures do not establish parallel trends; they show only that the pre-period data do not contradict the mixed model. No post-2013 coefficient is distinguishable from zero under either model. Section~\ref{sec:sim-panel} shows what the two-way intervals miss when the fitted GP--OU model is correct, and how the mixed model behaves when it is not; neither simulation shows which model is right for these data. Neither analysis indicates an effect of expansion on P\&S syphilis rates.

\subsection{Functional associations with dependence-aware inference}
All analyses in this section use the prespecified full covariate set; no selected adjustment set is used. For the level and timing of 2014--2019 incidence, the covariates are the 2008--2013 level and trend of P\&S rates, the 2008--2013 gonorrhea rate, percentages Black and in poverty, and latitude. Propensities are clipped at $0.02$ and $0.98$ (three states), so the estimates are clipped AIPW associations. They estimate the ATE (Theorem~\ref{thm:causalid}) if Assumption~\ref{ass:unconf} holds with this set and, because of the clipping, if in addition either the outcome regressions are correct or the propensity model is correct and no true propensity lies outside $[0.02,0.98]$ (Section~\ref{sec:ate}); adoption may also depend on unmeasured political factors. As an independence-based diagnostic, the unit-level fractional-sampling collection at nominal level $0.90$ retains all 64 candidate sets. It has no demonstrated coverage guarantee under spatial dependence, and with 44 states it does not distinguish sufficient sets in any case. Table~\ref{tab:functional} reports the estimates with standard errors from the stacked influence function under each covariance estimator, and Figure~S1 shows the estimated timing association map.

\begin{table}[t]\centering\footnotesize
\caption{Functional AIPW associations of the 2014 Medicaid expansion cohort with the 44 jurisdictions and prespecified full adjustment: estimates and standard errors from the stacked influence function under different covariance estimators (Mehler and chord use the 260 km range of the fitted latent Gaussian model; blocks are Census divisions). Level of log cumulative incidence, and mean timing in years.}\label{tab:functional}
\begin{tabular}{llrr}\toprule
Analysis & Covariance & Level & Mean timing\\\midrule
States, 2014--2019 & i.i.d. & 0.381 (0.379) & 0.074 (0.156)\\
 & HAC & 0.381 (0.374) & 0.074 (0.188)\\
 & Block & 0.381 (0.375) & 0.074 (0.221)\\
 & Mehler & 0.381 (0.453) & 0.074 (0.234)\\
 & Chord & 0.381 (0.559) & 0.074 (0.238)\\
 & Full-refit bootstrap & 0.381 (0.968) & 0.074 (0.336)\\
\midrule
States, placebo 2008--2013 & i.i.d. & 0.314 (0.168) & -0.036 (0.163)\\
 & HAC & 0.314 (0.159) & -0.036 (0.151)\\
 & Block & 0.314 (0.104) & -0.036 (0.138)\\
 & Mehler & 0.314 (0.178) & -0.036 (0.173)\\
 & Chord & 0.314 (0.251) & -0.036 (0.251)\\
 & Full-refit bootstrap & 0.314 (0.445) & -0.036 (0.601)\\
\bottomrule\end{tabular}\end{table}

The level association is $0.38$ and the mean-timing association $0.07$ years; neither is significant under any estimator. The Mehler and chord standard errors of the level are 1.2 and 1.5 times the independence, HAC and block values, and for timing about 1.5 times the independence value. The $L^2$ test of no timing effect has $p=0.73$, $0.84$ and $0.91$ with the independence, Mehler and chord covariances. The placebo shows why the covariance matters. For 2008--2013 outcomes with 2002--2007 covariates, the level association is $0.31$. Its interval excludes zero with Census-division blocks ($0.11$ to $0.52$) and barely with HAC, barely includes zero with the Mehler covariance ($-0.03$ to $0.66$) and clearly includes it with the chord covariance ($-0.18$ to $0.81$). Whether a spurious pre-period ``effect'' is declared thus depends on how dependence is handled; with 44 units in nine divisions, block and kernel estimators underestimate the variance (Section~\ref{sec:sim-dependence}). Clipping propensities at 0.05 instead of 0.02 lowers the level association to 0.31 with smaller standard errors, so that its independence interval excludes zero, and halving or doubling the working range changes the chord standard error by $-22\%$ and $+40\%$ (Section~S6); conclusions based on the Mehler and chord covariances do not change. A full-refit bootstrap of states is unstable, with a level standard error 2.5 times the independence value, because some resamples nearly separate the exposure groups.

\paragraph{Counties} For 2,144 counties in the 36 states that expanded in 2014 or had not expanded by 2023, the 2023 rate among women aged 15--44 gives a scalar outcome (Section~S6). Its propensities are fitted to county covariates, although the policy is assigned to states, so a county-level contrast is a statistical functional rather than the effect of a feasible intervention; Section~\ref{sec:app-causal} defines a state-level intervention. The clipped AIPW association with $\log(1+\text{rate})$ is $0.07$. Its standard error is $0.17$ under independence, $0.28$ with states, the units of assignment, as blocks, $0.31$ with a state-cluster refit bootstrap, and $0.67$ with the chord estimator, using the 429 km range of a REML Gaussian-process fit; no estimator indicates an association. The chord bound is loose here because counties are dense relative to the range. The point estimate is fragile: it moves from $-0.15$ without clipping to $0.27$ with clipping at 0.05, and a Gaussian-process regression with linear covariate adjustment gives a significant coefficient, $0.55$ (SE $0.17$). The county analysis is therefore model dependent and descriptive.

\subsection{Stochastic interventions and mediation}\label{sec:app-causal}
Table~\ref{tab:causal} reports two further estimands. Incremental interventions halve or double each state's odds of expanding in 2014. The estimated state propensities range from $0.014$ to $0.988$. The AIPW estimates were stabilized by clipping, whereas incremental effects need neither clipping nor positivity (Theorem~\ref{thm:causalid}), at the price of a correctly specified propensity model. For states, the effects on the level are $-0.005$ ($\delta=1/2$) and $-0.002$ ($\delta=2$), with chord standard errors $0.014$ and $0.016$, and the effects on mean timing are at most $0.01$ years in absolute value. They are small because most propensities are near zero or one, so rescaling the odds changes few exposure probabilities materially.

For counties the intervention acts where the policy is assigned. A logistic model for state adoption uses the state covariates above. Each state's odds are multiplied by $\delta$, and the estimand is the change in the county-average outcome relative to the observed assignment, with outcome regressions on county and state covariates. Unconfoundedness is then required at the state level: given the state covariates, adoption is independent of the county covariates and potential outcomes. The stacked estimating equations are summed within states, so the influence function is a state-level quantity and the 36 states are the units of inference. The state-cluster standard errors treat states as independent. As a working-model sensitivity analysis, the Mehler and chord estimators are also applied to the state aggregates with state centroids and the larger of the two fitted ranges (429 km); they carry no formal conservativeness guarantee here. The state aggregates are sums over different numbers of counties, so they are not identically distributed, and aggregation over different county configurations need not preserve an exponential covariance in centroid distance; the theory of Section~\ref{sec:dependence}, including Theorem~\ref{thm:covest}(d), therefore does not apply. In addition, the fitted state propensities range from $0.006$ to $0.9998$, close to separation with 36 states and six covariates, so the required correct propensity model is fragile. The estimand weights counties equally, so states with many counties weigh more. Doubling every state's odds changes the county mean of $\log(1+\text{rate})$ by $0.041$ and halving them by $-0.037$, with state-cluster standard errors $0.031$ and $0.029$ and chord standard errors $0.044$ and $0.042$. Neither is distinguishable from zero.

\begin{table}[t]\centering\footnotesize
\caption{Stochastic-intervention and mediation effects of the 2014 Medicaid expansion: estimates for the level (log total rate; for counties $\log(1+\text{rate})$) and the mean timing (years), with independence (I), Mehler (M) and chord (C) standard errors. Incremental effects compare multiplying every state's odds of expansion by $\delta$ with the observed assignment. $^a$State-level intervention, county-average outcome, influence functions summed within the 36 states; its I column treats states as independent (state-cluster standard errors). $^b$Mediator: chlamydia rate in the first three years (2014--2016; placebo 2008--2010); outcome: P\&S syphilis in the last three (2017--2019; placebo 2011--2013); propensities clipped at 0.02, so these are clipped-AIPW-type functionals (Section~\ref{sec:app-causal}).}\label{tab:causal}
\begin{tabular}{llrrrrrrrr}\toprule
 & & \multicolumn{4}{c}{Level} & \multicolumn{4}{c}{Mean timing}\\\cmidrule(lr){3-6}\cmidrule(lr){7-10}
Analysis & Effect & Est. & I & M & C & Est. & I & M & C\\\midrule
States, 2014--2019 & Incremental, $\delta=1/2$ & -0.005 & 0.009 & 0.011 & 0.014 & -0.000 & 0.009 & 0.010 & 0.014\\
 & Incremental, $\delta=2$ & -0.002 & 0.011 & 0.014 & 0.016 & -0.010 & 0.016 & 0.017 & 0.021\\
\midrule
Placebo, 2008--2013 & Incremental, $\delta=1/2$ & -0.028 & 0.014 & 0.015 & 0.020 & 0.001 & 0.022 & 0.023 & 0.030\\
 & Incremental, $\delta=2$ & 0.020 & 0.018 & 0.019 & 0.028 & 0.005 & 0.023 & 0.023 & 0.032\\
\midrule
Counties, 2023$^a$ & Incremental, $\delta=1/2$ & -0.037 & 0.029 & 0.038 & 0.042 & \multicolumn{4}{c}{--}\\
 & Incremental, $\delta=2$ & 0.041 & 0.031 & 0.040 & 0.044 & \multicolumn{4}{c}{--}\\
\midrule
Mediation, 2017--2019$^b$ & Total, clipped functional & 0.210 & 0.471 & 0.500 & 0.633 & 0.018 & 0.039 & 0.043 & 0.100\\
 & NDE functional & 0.251 & 0.548 & 0.586 & 0.793 & 0.008 & 0.039 & 0.066 & 0.094\\
 & NIE functional & -0.041 & 0.088 & 0.111 & 0.176 & 0.009 & 0.011 & 0.011 & 0.018\\
\midrule
Placebo, 2011--2013$^b$ & Total, clipped functional & 0.208 & 0.250 & 0.280 & 0.362 & 0.036 & 0.207 & 0.218 & 0.348\\
 & NDE functional & 0.166 & 0.252 & 0.282 & 0.401 & 0.035 & 0.210 & 0.229 & 0.353\\
 & NIE functional & 0.041 & 0.079 & 0.080 & 0.151 & 0.001 & 0.010 & 0.010 & 0.015\\
\bottomrule\end{tabular}\end{table}

For mediation, the windows are ordered in time. The mediator is the logarithm of the 2014--2016 chlamydia rate; the outcome is the level and mean timing of 2017--2019 P\&S syphilis; the covariates are from 2008--2013 and include the chlamydia rate in place of gonorrhoea. The placebo shifts all windows back by six years. We use a Gaussian linear mediator model and outcome regressions on $(W,X,M)$ with slopes common to the two arms, which suits 44 units, and propensities clipped at $0.02$. This analysis is a descriptive illustration of the mediation functionals, not evidence about a screening mechanism, for two reasons. First, reported chlamydia reflects both screening and incidence. Second, sequential ignorability is strong: shared sexual-network risk is a plausible mediator--outcome confounder, captured only partly by the pre-period levels through the persistence of the latent field.

Because the propensities are clipped, the mediation score remains consistent for the natural effects only on the branch where the mediator and outcome models are both correct; the two branches that rely on the propensity are lost wherever clipping is active, so Table~\ref{tab:causal} reports clipped functionals. The estimated exposure--mediator contrast is small (coefficient $-0.013$ on the log scale, chord standard error $0.029$). The indirect-effect functional is $-0.04$ for the level (chord standard error $0.18$) and $0.01$ years for mean timing ($0.02$), and the direct-effect functional for the level is $0.25$ ($0.79$). In the placebo period the indirect-effect functional for the level is $0.04$ ($0.15$).

An illustrative coefficient-bias sensitivity calculation gauges mediator--outcome confounding. In the linear working models the indirect effect is approximately $\alpha_W\gamma_M$. Here $\alpha_W$ is the exposure coefficient of the mediator model, which such confounding does not bias under Assumption~\ref{ass:seqign}(i), and $\gamma_M$ is the mediator coefficient of the outcome model, estimated as $2.2$ for the level. A confounding bias $b$ in $\gamma_M$ changes the indirect effect by $-\alpha_Wb$. Taking $|\alpha_W|=0.07$, the upper end of its nominal 95\% chord interval, a bias as large as twice $\widehat\gamma_M$ changes the product by $0.31$, comparable to the sampling uncertainty of the indirect effect on the level. This is not a formal bound for the reported clipped, augmented estimator: that estimator is not exactly the coefficient product, the interval endpoint gives no simultaneous confidence statement, and the calculation assumes that the exposure--mediator contrast remains identified. The data are thus uninformative about mediation through chlamydia-based screening rather than evidence against it. Throughout Table~\ref{tab:causal} the chord standard errors are 1.3 to 2.5 times the independence values for states, and about 1.45 times the state-cluster values for the county effects.

\section{Comparison with point-process spatio-temporal causal inference}\label{sec:geocausal}
The closest methodological alternative for spatio-temporal causal questions is the framework of \citet{papadogeorgou2022causal}, extended to heterogeneous effects by \citet{zhou2024heterogeneous} and to mediation by \citet{mukaigawara2025spatiotemporal}, and implemented in the R package \texttt{geocausal} \citep{mukaigawara2024geocausal}. There the treatment in each period is a spatial point pattern on a region $\Omega_0$, such as the locations of airstrikes, and the estimand is the expected number of outcome events in a set $B\subseteq\Omega_0$ when the treatment patterns of the last $L$ periods follow a stochastic intervention. Table~\ref{tab:geocausal} compares the two approaches.

\begin{table}[t]\centering\footnotesize
\caption{The point-process framework implemented in \texttt{geocausal} and the present approach.}\label{tab:geocausal}
\begin{tabular}{>{\raggedright\arraybackslash}p{2.6cm}>{\raggedright\arraybackslash}p{5.95cm}>{\raggedright\arraybackslash}p{5.95cm}}\toprule
 & \texttt{geocausal} \citep{papadogeorgou2022causal,mukaigawara2025spatiotemporal} & This article\\\midrule
Treatment & Point pattern per period on a continuous region & Categorical exposure of each areal unit\\
Outcome & Point pattern; event counts in regions $B$ & Distribution-valued, embedded in a Hilbert space (e.g.\ level and timing law)\\
Interference & Arbitrary spatial spillover and temporal carryover & None; dependence through a latent Gaussian field\\
Estimands & Stochastic-intervention effects over $L$ periods; heterogeneous effects; interventional mediation effects & ATE; stochastic and incremental effects; natural direct and indirect effects\\
Unconfoundedness & Treatment pattern independent of all past and future potential outcomes given the observed history & Unit-level, given a sufficient set $X_S$ (Assumption~\ref{ass:unconf})\\
Overlap & Bounded relative overlap $e\ge c_0f$ & Weighted moment condition (Assumption~\ref{ass:overlap}); automatic for incremental interventions\\
Adjustment & Propensity model (covariates, lags) specified by the analyst & Exhaustive collection of sufficient sets (independent data)\\
Estimators & Spatially smoothed inverse probability weighting, with H\'ajek versions \citep{zhou2024heterogeneous}; correct propensity model required & Doubly (ATE) and triply (mediation) robust scores with stacked nuisances\\
Asymptotics & Number of periods $T\to\infty$; martingale arguments & Number of units $n\to\infty$ (increasing domain); $\rho$-mixing of the latent field\\
Variance & Not identified; consistently estimable upper bound & Long-run covariance: consistent Mehler or conservative chord estimator\\
Application & Daily airstrikes and insurgent attacks, Iraq & Annual syphilis surveillance and a one-time state policy\\\bottomrule
\end{tabular}
\end{table}

\paragraph{Theory and applications} \texttt{geocausal} obtains replication from many periods over one region, leaves interference unrestricted and requires a correct treatment model, with a conservative variance bound because the variance is not identified. The present approach obtains replication from units, rules out interference and models dependence through a latent field. In the Iraq application, strikes recur daily and spill over kilometres, so interference is the substance of the question and our assumptions would be untenable. Medicaid expansion is a one-time areal adoption with essentially one assignment per state, so replication must come from jurisdictions, and the outcome of interest is a distribution over time. A natural combination would take the kernel-smoothed outcome surface of \texttt{geocausal}, an element of $L^2(\Omega_0)$, as the Hilbert-valued outcome of Section~\ref{sec:effects}.

\section{Discussion}
The results connect subspace asymptotics to a basis-invariant adjustment tensor, a calibrated simultaneous outer collection and functional causal contrasts, and extend the inferential part to spatially and temporally dependent units. Two points may be of independent interest. For an asymptotically linear statistic, fractional sampling with the finite-population factor is as easy to justify as subsampling and behaves far better in moderate samples. And under a latent Gaussian model the covariance of any square-integrable score is a power series in the latent correlation with positive semidefinite coefficients. This yields an exact undercoverage formula for independence-based intervals, a parametric covariance estimator, and a convex upper bound that remains valid whatever the nonlinearity of the score and whenever the spatial range is overstated. The same latent structure underlies Gaussian Markov random field and Gaussian-process disease models, so the range can be borrowed from the outcome model that practitioners already fit. Because the ATE, incremental-intervention and mediation estimators are all solutions of stacked estimating equations, one central limit theorem and one pair of covariance estimators serve every causal estimand in the paper.

Several boundaries remain. The number of covariates is fixed, exhaustive evaluation is exponential in $p$, and structural dimensions are supplied. The root-$n$ projector expansion is proved for independent data; under dependence, linearization of the criterion tensor is an assumption of Theorem~\ref{thm:blockhalf}, so the selection procedure as a whole is established only for independent data, and in the application the collection served as an independence-based diagnostic while the prespecified full set was used. The covariance theorem covers fixed distance bins and finite chaos order; the adaptive binning of our default chord implementation and the Mehler estimator with $m_0=2$ for nonlinear scores are supported numerically, not by the theorem. The dependence model assumes a common latent correlation across components, identically distributed units, increasing-domain sampling and no interference, so it is a working model when covariates such as latitude are fixed site characteristics. The covariance guarantees are asymptotic: with few units per correlation range, the Mehler and even the chord intervals can undercover, and a generous range is the safer choice. The difference-in-differences simulations show how each method behaves when the GP--OU model is or is not correct, not which model describes the data. In the surveillance analysis, the county association changes sign with the propensity clipping threshold, the two-way and mixed-model estimates differ in sign, and the mediation analysis rests on a proxy mediator and strong ignorability assumptions, so all observational comparisons remain exploratory. The causal conclusions rest on unconfoundedness given the prespecified covariates and, for mediation, on a cross-world sequential ignorability condition; with exposure-related unmeasured spatial confounding or mediator--outcome confounding, these estimands are not identified, and sensitivity analysis \citep{imai2010general,tchetgen2012semiparametric} would be the next step. Natural extensions are a dependent-data version of Theorem~\ref{thm:rate}, a finite-sample correction for the covariance estimators, and designs with interference. For the last, the stochastic-intervention estimands of \texttt{geocausal} (Section~\ref{sec:geocausal}), combined with distribution-valued outcomes and unit-level replication, are a promising route.

\section*{CRediT authorship contribution statement}
\textbf{Hsin-Hsiung Huang:} Conceptualization, Methodology, Formal analysis, Software, Data curation, Visualization, Writing -- original draft, Writing -- review \& editing, Funding acquisition.

\section*{Declaration of competing interest}
The author declares that he has no known competing financial interests or personal relationships that could have appeared to influence the work reported in this paper.

\section*{Funding}
This work was supported in part by the National Science Foundation [grant numbers DMS-1924792, DMS-2318925]; and the National Institute of Allergy and Infectious Diseases of the National Institutes of Health [grant number 1R01AI206603-01].

\section*{Data availability}
All data are public: CDC AtlasPlus surveillance exports, Medicaid expansion dates from published replication materials, and federal statistical covariates. The data sources, with URLs and the construction of the analysis files, are listed in Section~S4 of the Supplementary Material. The code that reproduces the analyses is available from the author upon request.

\appendix
\section{Proofs}\label{app:proofs}
\subsection{Identification and overlap}
\begin{proof}[Proof of Theorem~\ref{thm:causalid}]
Fix $w$ with $P\{f(w\mid X_S)>0\}>0$. On $\{f(w\mid X_S)>0\}$ we have $e(w\mid X_S)>0$, so $m_w(X_S)=\E(\widetilde Y\mid X_S,W=w)$ is defined there. By consistency and Assumption~\ref{ass:unconf}, $\E(\widetilde Y\mid X_S,W=w)=\E\{\widetilde Y(w)\mid X_S,W=w\}=\E\{\widetilde Y(w)\mid X_S\}$. Bochner integrability follows from $\E\|\widetilde Y(w)\|<\infty$. Because $f(w\mid X_S)$ is $X_S$-measurable, the tower property gives $\E\{f(w\mid X_S)\widetilde Y(w)\}=\E\{f(w\mid X_S)m_w(X_S)\}$, and summing over the finitely many $w$ gives the first identity in \eqref{eq:gformula}. For the second,
\[
\E\Bigl\{\frac{f(W\mid X_S)}{e(W\mid X_S)}\widetilde Y\Bigm|X_S\Bigr\}=\sum_{w:e(w\mid X_S)>0}e(w\mid X_S)\frac{f(w\mid X_S)}{e(w\mid X_S)}m_w(X_S)=\sum_wf(w\mid X_S)m_w(X_S),
\]
where terms with $e(w\mid X_S)=0$ have $f(w\mid X_S)=0$ by Assumption~\ref{ass:overlap}(a).

For the score, let $\widetilde m_w$ and $\widetilde e$ be working functions with $\widetilde e(w\mid X_S)>0$ where $f(w\mid X_S)>0$. Conditioning on $X_S$,
\[
\E(\widetilde\psi_F\mid X_S)=\sum_wf(w\mid X_S)\Bigl[\widetilde m_w+\frac{e(w\mid X_S)}{\widetilde e(w\mid X_S)}\{m_w-\widetilde m_w\}\Bigr](X_S),
\]
which equals $\sum_wfm_w$ if $\widetilde m_w=m_w$ or $\widetilde e=e$; taking expectations proves $\E\widetilde\psi_F=\mu(F)$. For square integrability with the true nuisances, $\E\|\psi_F\|^2\le2\E\|\sum_wfm_w\|^2+2\E\{f(W\mid X_S)^2\|\widetilde Y-m_W\|^2/e(W\mid X_S)^2\}$. The first term is finite by Jensen's inequality and the second moments, and the second equals $2\E\{\sum_wf(w\mid X_S)^2\sigma_w^2(X_S)/e(w\mid X_S)\}<\infty$ by Assumption~\ref{ass:overlap}(b).

For the incremental intervention write $e=e(X_S)$ and $D=\delta e+1-e>0$. Then $f_\delta(1\mid x)/e=\delta/D$ and $f_\delta(0\mid x)/(1-e)=1/D$. If $\delta\ge1$, then $D\ge1$, so the ratios are at most $\delta$ and $1$; if $\delta<1$, then $D\ge\delta$, so they are at most $1$ and $\delta^{-1}$. Hence $f_\delta/e\le\max(\delta,\delta^{-1})$, which is the stated inequality; it implies (a), and (b) with constant $\max(\delta,\delta^{-1})\sum_w\E\{f_\delta(w\mid X_S)\sigma_w^2(X_S)\}\le\max(\delta,\delta^{-1})\sum_w\E\sigma_w^2(X_S)$.
\end{proof}

\begin{proof}[Proof of Theorem~\ref{thm:mediation}]
(a) Write $x$ for $X_S$ and $m$ for a mediator value. By the tower property,
\[
\theta_{ab}=\E\Bigl[\int\E\{\widetilde Y(a,m)\mid X_S,M(b)=m\}\,P\{M(b)\in dm\mid X_S\}\Bigr].
\]
By (i), $P\{M(b)\in\cdot\mid X_S\}=P\{M(b)\in\cdot\mid X_S,W=b\}=P_{M\mid X_S,b}$, and
\begin{align*}
\E\{\widetilde Y(a,m)\mid X_S,M(b)=m\}&=\E\{\widetilde Y(a,m)\mid X_S,W=b,M(b)=m\}\\
&=\E\{\widetilde Y(a,m)\mid X_S,W=b\}=\E\{\widetilde Y(a,m)\mid X_S\},
\end{align*}
using (i), then (ii), then (i). By (i) again, then (ii) with $b=a$, and consistency, the last expression equals $\E\{\widetilde Y(a,m)\mid X_S,W=a,M(a)=m\}=\mu_a(X_S,m)$ for $P_{M\mid X_S,a}$-almost every $m$, and hence for $P_{M\mid X_S,b}$-almost every $m$ by the absolute continuity in (iii). This gives $\theta_{ab}=\E\eta_{ab}(X_S)$. The expression for $r_{ab}$ is Bayes' formula: both conditional laws have densities $P(W=w\mid x,m)/e(w\mid x)$ with respect to the law of $M$ given $X_S=x$.

(b) Condition on $X_S$ throughout and write $\E_w$ for expectation given $X_S$ and $W=w$. Since $\E\{\I(W=w)Z\mid X_S\}=e(w\mid X_S)\E_wZ$,
\begin{align*}
\E(\widetilde\psi_{ab}\mid X_S)={}&\frac{e(a\mid X_S)}{\widetilde e(a\mid X_S)}\E_a\bigl[\widetilde r_{ab}\{\mu_a-\widetilde\mu_a\}(X_S,M)\bigr]\\
&+\frac{e(b\mid X_S)}{\widetilde e(b\mid X_S)}\bigl\{\E_b\widetilde\mu_a(X_S,M)-\widetilde\eta_{ab}(X_S)\bigr\}+\widetilde\eta_{ab}(X_S),
\end{align*}
where the first term uses $\E_a\{\widetilde Y\mid M\}=\mu_a(X_S,M)$. (1) If the outcome regression and the mediator law are correct, the first term vanishes and $\widetilde\eta_{ab}=\E_b\mu_a(X_S,M)=\eta_{ab}$, so the second vanishes and the sum is $\eta_{ab}$. (2) If the propensity and the mediator law are correct, then $\widetilde r_{ab}=r_{ab}$ and the first term is $\E_b\{\mu_a-\widetilde\mu_a\}(X_S,M)=\eta_{ab}-\widetilde\eta_{ab}$, because $\widetilde\eta_{ab}=\E_b\widetilde\mu_a$ under the correct mediator law; the second term vanishes and the sum is $\eta_{ab}$. (3) If the propensity and the outcome regression are correct, the first term vanishes and the last two give $\E_b\mu_a(X_S,M)=\eta_{ab}$. In each case $\E\widetilde\psi_{ab}=\E\eta_{ab}(X_S)=\theta_{ab}$ by (a).

(c) With $a=b$, $r_{aa}=1$ and $\eta_{aa}=m_a$, so $\psi_{aa}=m_a+\I(W=a)(\widetilde Y-m_a)/e(a\mid X_S)$. For scalar outcomes, $\psi_{ab}-\theta_{ab}$ is the efficient influence function derived by \citet{tchetgen2012semiparametric}. For Hilbert-valued outcomes, apply that result to $\langle h,\widetilde Y\rangle$ for each $h\in\Hcal$: the functional is linear in the outcome and the score is linear in $\widetilde Y$, $\mu_a$ and $\eta_{ab}$.
\end{proof}

\begin{proof}[Proof of Theorem~\ref{thm:ident}]
Strict positivity makes $P_X$ and $P_{X\mid W=w}$ mutually absolutely continuous. Ignorability and consistency identify the same conditional response kernel $x\mapsto\mathcal L\{Y(w)\mid X=x\}=\mathcal L(Y\mid X=x,W=w)$ under these equivalent covariate laws. A subspace is sufficient exactly when this kernel is a measurable function of its projection, so the observed and potential-outcome central subspaces coincide.

Fix $w,A$, and put $U=B_w\tp X$, $V=B_W\tp X$, $Z=X_A$. For bounded $h,k$, ignorability and iterated conditional expectation give
\[
\Cov\{h(Y(w)),k(W)\mid Z\}=\Cov\{g_h(U),g_k(V)\mid Z\},
\]
where $g_h(U)=\E\{h(Y(w))\mid X\}$ and $g_k(V)=\E\{k(W)\mid X\}$. Conditional independence of $U,V$ makes every covariance vanish; detectability gives the converse. Therefore $A$ is sufficient if and only if $U\indep V\mid Z$.

Since $W\indep X\mid V$, we have $W\indep U\mid(V,Z)$. Together with $U\indep V\mid Z$, contraction and weak union imply $U\indep V\mid(Z,W)$. Conversely, suppose $U\indep V\mid(Z,W)$. For bounded $f$, versions of the conditional expectations satisfy
\[
a(V,Z):=\E\{f(U)\mid V,Z\}=\E\{f(U)\mid V,Z,W\}=b(Z,W)
\]
almost surely. For fixed $Z=z$, the conditional laws of $V$ given $(Z=z,W=v)$ are mutually absolutely continuous: positivity preserves equivalence after conditioning, or, directly, their Gaussian supports are the same affine subspace
$B_{W,A}\tp z+\operatorname{range}(B_{W,-A}\tp)$, with strictly positive densities there. Hence $a(\cdot,z)$ equals the same constant $b(z,v)$ under every arm; it is independent of $V$. Thus $U\indep V\mid Z$.

Given $(Z,W=v)$, $(U,V)$ is jointly Gaussian, possibly degenerate, with cross covariance
\[B_w\tp H_v(A)B_W.\]
Gaussian independence is equivalent to zero cross covariance, including degenerate blocks. Summing its nonnegative operator norms proves the equivalence in \eqref{eq:criterion}. The projector formula proves invariance to the choice of bases.
\end{proof}

\begin{proof}[Proof of Proposition~\ref{prop:overlap}]
Let $q_v=P(W=v)>0$ and $f_v$ be the Gaussian density of $X\mid W=v$. Bayes' rule gives $f_v(x)/f_u(x)=(q_u/q_v)\pi_v(x)/\pi_u(x)$. A uniform positive lower bound for every propensity bounds this ratio above and away from zero. The log ratio is a polynomial of degree at most two in $x$. A nonconstant quadratic or linear polynomial cannot be bounded both above and below on $\R^p$. Equivalence with Lebesgue measure therefore forces its quadratic and linear coefficients to vanish. Thus the covariance matrices and means coincide, and the treatment is independent of $X$.

With a common covariance, $f_u/f_v$ is the exponential of an affine function of $X$. Since $\pi_v^{-1}=\sum_u q_u f_u/(q_v f_v)$, its $r$th power is bounded by a finite constant times the sum of these exponential-affine terms to the $r$th power. All have finite expectation under every Gaussian component, hence under their finite mixture. Positivity follows because all densities are positive.
\end{proof}

\begin{proof}[Proof of Proposition~\ref{prop:kernel}]
If $\phi(y)=\phi(y')$, the kernel embeddings of the two Dirac probability measures coincide. Characteristicness gives $y=y'$. Continuity of the kernel makes $\phi$ continuous; its image is contained in a separable Hilbert space, since the domain is separable and the span of the image of a countable dense subset is dense in the feature span. The Lusin--Souslin theorem \citep{kechris1995classical} gives a Borel inverse on the image. Hence $\sigma\{\phi(Y)\}=\sigma(Y)$ and all conditional-independence statements, including those defining central subspaces, are preserved. Separable Hilbert norm distance and powered Euclidean distance for $0<\alpha<2$ have strong negative type \citep{lyons2013distance}. Their distance covariance characterizes independence when first distance moments are finite. The feature distance is bounded and the stated predictor moment supplies that integrability.
\end{proof}

\subsection{The subspace expansion}\label{app:rate}
We give the empirical-process details to distinguish the rate conclusion from a mere uniform law of large numbers. The parameter spaces below are finite dimensional and compact after restricting $R$ to a neighborhood of $I$.

\paragraph{Kernel classes}
For symmetric $M$, $u\mapsto u\tp Mu$ lies in a finite-dimensional vector space, and is VC-subgraph. Composition with $t\mapsto t_+^{\alpha/2}$, multiplication by a fixed nonnegative response-distance function, and finite sums give polynomial covering bounds with an envelope for the classes $\{h_D\}$ and $\{h_D-h_{D'}\}$. One can take
\[
F(o_1,\ldots,o_4)=C\sum_{j\ne k}\|z_j-z_k\|^\alpha\sum_{l\ne m}\rho(v_l,v_m).
\]
For the distances considered, $\rho(v_l,v_m)\le C\{\rho(v_l,v_0)+\rho(v_m,v_0)\}$ (with an innocuous additive constant for bounded distances). Cauchy--Schwarz and the fourth-power moment assumptions give $\E F^2<\infty$, including terms sharing an observation. Hoeffding projections preserve the polynomial entropy bounds. Standard maximal inequalities for canonical U-processes \citep{nolan1987u,sherman1994maximal} then give, for order $k\ge2$,
\begin{equation}\label{eq:upro}
\sup_f|U_{n,k}f|=O_P(n^{-k/2}),\qquad
\sup_{f\in\mathcal F_n}|U_{n,k}f|=o_P(n^{-k/2})
\end{equation}
when the localized class has $\sup_{f\in\mathcal F_n}P^k f^2\to0$. The second conclusion follows by the localized entropy bound and its vanishing $L^2$ radius. These are statements for fully canonical kernels under independent sampling.

\begin{proof}[Proof of Theorem~\ref{thm:rate}]
The covariance estimator satisfies
\[
\widehat\Sigma_X-\Sigma_X=(\Pn-P)\{(X-\mu)(X-\mu)\tp\}+O_P(n^{-1}).
\]
Differentiability of the inverse square root on positive-definite matrices yields
\begin{equation}\label{eq:R}
\widehat R-I=\Pn\mathcal L_\Sigma[(X-\mu)(X-\mu)\tp-\Sigma_X]+O_P(n^{-1}).
\end{equation}
Derivatives in the chart are understood after vectorizing $\theta$. Sample centering has no additional effect because the objective uses pairwise predictor differences. In fact $\widehat Z_i-\widehat Z_j=\widehat R(Z_i-Z_j)$ exactly.

The polynomial entropy bound and integrable envelope give a uniform U-process LLN. Continuity of $\Gamma(R\tp C)$ and compactness imply uniform convergence after substituting $\widehat R\to I$. Uniqueness modulo rotation separates the maximizing subspace from every closed set outside a fixed projector neighborhood. The global near-maximizer condition and the argmax argument give consistency, so its chart coordinate $\widehat\theta$ exists with probability tending to one and converges to zero.

Write $G=Ph$ in chart coordinates. By Hoeffding decomposition,
\begin{align*}
G_n(\theta,R)-G_n(0,R)
={}&G(\theta,R)-G(0,R)\\
&+4(\Pn-P)\{\tau(\cdot;\theta,R)-\tau(\cdot;0,R)\}
+\sum_{k=2}^4 \binom{4}{k}U_{n,k}\pi_k(h_{R\tp C(\theta)}-h_{R\tp C(0)}).
\end{align*}
The first derivative $\partial_\theta\tau(\cdot;0,R)$ is Lipschitz in $R$ with square-integrable envelope by the second-derivative bound, hence Donsker. The continuous second-derivative class with integrable envelope on the compact neighborhood is Glivenko--Cantelli. Using the integral form of Taylor's formula, its centered quadratic remainder is $o_P(\|\theta\|^2)$ uniformly near zero. Thus the linear-projection term is
$O_P(n^{-1/2}\|\theta\|)+o_P(\|\theta\|^2)$.
The degenerate terms are $O_P(n^{-1})$ uniformly by \eqref{eq:upro}. The mixed population increment is bounded by
$C\|\theta\|\|\widehat R-I\|=O_P(n^{-1/2}\|\theta\|)$.
Negative curvature and comparison with the objective at zero therefore give
\[
\kappa\|\widehat\theta\|^2\le O_P(n^{-1/2}\|\widehat\theta\|)+o_P(\|\widehat\theta\|^2)+O_P(n^{-1}),
\]
which implies $\widehat\theta=O_P(n^{-1/2})$ by absorbing the quadratic remainder and solving the resulting quadratic inequality.

For $\|\theta\|\le M/\sqrt n$ and $\|R-I\|\le M/\sqrt n$, dominated convergence makes the $L^2$ diameter of the kernel-increment class tend to zero. The order-two term in \eqref{eq:upro} is now $o_P(n^{-1})$; orders three and four are already smaller. Taylor expansion of the population term, stochastic equicontinuity of the first-derivative class, and \eqref{eq:R} give uniformly on this neighborhood
\[
G_n(\theta,\widehat R)-G_n(0,\widehat R)
=\tfrac12\theta\tp\mathbb V\theta+\theta\tp\Pn\xi+o_P(n^{-1}).
\]
The quadratic maximizer is $-\mathbb V^{-1}\Pn\xi$. Comparing it with the near-maximizer and using negative definiteness shows that their distance is $o_P(n^{-1/2})$, proving \eqref{eq:bahadur}. This is the local quadratic U-process argument underlying \citet{sherman1993limiting}; the mixed whitening term is explicit here.

In standardized coordinates the estimated directions are
$\widehat D=\widehat R\tp C(\widehat\theta)$, and
\[
\widehat D-C_0=C_{0\perp}\widehat\theta+(\widehat R-I)\tp C_0+O_P(n^{-1}).
\]
The derivative of $D\mapsto P_D$ at orthonormal $C_0$ is
$E\mapsto(I-P_{C_0})EC_0\tp+C_0E\tp(I-P_{C_0})$.
The original-coordinate map is $D\mapsto P_{\Sigma_X^{-1/2}D}$, also differentiable at full-column-rank $C_0$. Applying these derivatives proves the projector expansion with a square-integrable influence function. For arm fits, the conditional empirical measure has first-order increment
$\Pn\{\I(W=w)(f-P_wf)/q_w\}$ with $q_w=P(W=w)>0$. Applying the same expansion to each of finitely many fits gives their joint asymptotic linearity and CLT.
\end{proof}

\paragraph{Numerical smoothing bound}
For $0<\alpha<2$ and $t\ge0$, concavity gives
$0\le(t+\epsilon)^{\alpha/2}-t^{\alpha/2}\le\epsilon^{\alpha/2}$.
U-centering has a bounded row-sum operator norm, and $n^{-2}\sum_{ij}|\widetilde\rho_{ij}|=O_P(1)$ under the stated moments. Uniformly in $D$, the smoothed and unsmoothed objectives differ by $O_P(\epsilon_n^{\alpha/2})=O_P(n^{-2})$. A near-global maximizer of the smoothed objective with gap $o_P(n^{-1})$ is therefore also a near-global maximizer of the unsmoothed objective with that gap. This argument covers the smoothing change; it does not certify that a finite multistart optimizer attains the required gap.

\subsection{Collection recovery, subsampling and fractional sampling}
\begin{proof}[Proof of Theorem~\ref{thm:ridge}]
The map from projectors and covariance matrices to $\mathbf T$ is continuously differentiable wherever the required covariance blocks are positive definite. Theorem~\ref{thm:rate} and the covariance CLT therefore give joint asymptotic linearity and
$\max_{w,A}|\widehat Q_w(A)-Q_w(A)|=O_P(n^{-1/2})$, using the Lipschitz inequality for operator norms and finiteness of the indices.

Fix an arm. If there are insufficient candidates, their minimum criterion is a fixed $q_*>0$. With probability tending to one they precede every sufficient set in the decreasing ordering. At the true boundary, the ratio in \eqref{eq:ridge} tends to zero, because its numerator is $O_P(n^{-1/2})+c_n$ and its denominator is bounded away from zero. Ratios strictly before the boundary are bounded away from zero. Ratios after it tend to one, since $n^{-1/2}/c_n\to0$ in probability. The index-zero ratio is $c_0>0$, so the boundary is selected. When all candidates are sufficient, every nonzero-index ratio tends to one and index zero is selected. At least one sufficient candidate is needed for these cases to exhaust the possibilities; it is supplied by $[p]\in\Fcal$. A finite union over arms proves simultaneous recovery.
\end{proof}

\begin{proof}[Proof of Theorem~\ref{thm:subsample}]
Write $\psi_i=\psi_T(O_i)$, $\bar\psi=n^{-1}\sum_i\psi_i$ and $\Sigma_\psi=\Var\psi_T(O)$.

\emph{Case (i).} For a deterministic parameter vector $\mathbf T$, consider first the complete-subset empirical distribution of
$Z_I^0=\sqrt b\mathcal N(\widehat{\mathbf T}_I-\mathbf T)$. For a continuity point $t$ of $F(t)=P\{\mathcal N(Z)\le t\}$, its expectation equals
$P\{\sqrt b\mathcal N(\widehat{\mathbf T}_b-\mathbf T)\le t\}\to F(t)$, because a subset of size $b$ is an i.i.d.\ sample of that size. Its variance tends to zero even though the subsets overlap. Randomly permute the observations and split them into $\lfloor n/b\rfloor$ disjoint blocks of size $b$. The mean of the resulting bounded indicators has variance at most $1/(4\lfloor n/b\rfloor)$ by independence. Its conditional expectation given the unordered sample is the average over all size-$b$ subsets, so the conditional-variance inequality bounds the latter's variance by $O(b/n)$. Next, $s_{n,b}/\sqrt b\to1$ and
$s_{n,b}\|\widehat{\mathbf T}_n-\mathbf T\|=O_P(\sqrt{b/n})=o_P(1)$. Because $\mathcal N$ is Lipschitz, $|Z_I-Z_I^0|=o_P(1)$ uniformly over $I$. Sandwiching indicator functions at $t\pm\epsilon$ gives the same limit for the centred statistic.

\emph{Case (ii).} Let $I$ be a uniformly random $b$-subset independent of the data, and let $R_I=\widehat{\mathbf T}_I-\mathbf T-b^{-1}\sum_{i\in I}\psi_i$ and $R_n=\widehat{\mathbf T}_n-\mathbf T-\bar\psi$. Then
\[
s_{n,b}(\widehat{\mathbf T}_I-\widehat{\mathbf T}_n)=L_I+s_{n,b}(R_I-R_n),\qquad L_I=s_{n,b}\Bigl(\frac1b\sum_{i\in I}\psi_i-\bar\psi\Bigr).
\]
Unconditionally, $(O_i)_{i\in I}$ is an i.i.d.\ sample of size $b$, so $\sqrt b R_I\to0$ in probability by \eqref{eq:Tlinear}; also $\sqrt nR_n\to0$. Since $s_{n,b}^2=bn/(n-b)$ is of order $n$ and of order $b$, $P\{s_{n,b}\|R_I-R_n\|>\epsilon\}\to0$, the probability being over the data and $I$ jointly. Denote by $\pi_n(\epsilon)$ the conditional probability given the data. Then $\E\pi_n(\epsilon)\to0$, so $\pi_n(\epsilon)\to0$ in probability by Markov's inequality.

Conditionally on the data, $L_I$ is a normalized mean of a simple random sample without replacement from the finite population $\{\psi_1,\ldots,\psi_n\}$. It has conditional mean zero and conditional covariance
\[
s_{n,b}^2\,\frac1b\Bigl(1-\frac bn\Bigr)S_n=S_n,\qquad S_n=\frac1{n-1}\sum_{i=1}^n(\psi_i-\bar\psi)(\psi_i-\bar\psi)\tp,
\]
which converges almost surely to $\Sigma_\psi$. Fix $\lambda$ with $\lambda\tp\Sigma_\psi\lambda>0$ and put $a_i=\lambda\tp\psi_i$. The variance of the sampled sum is $b(1-b/n)\lambda\tp S_n\lambda$, which is of order $n$. Since $\E a^2<\infty$, $\max_{i\le n}a_i^2/n\to0$ almost surely. Hence for every $\epsilon>0$ the set $\{i:|a_i-\bar a|>\epsilon\{b(1-b/n)\lambda\tp S_n\lambda\}^{1/2}\}$ is eventually empty almost surely, which verifies H\'ajek's Lindeberg condition for sampling without replacement \citep{hajek1960limiting}. Thus $\lambda\tp L_I$ is conditionally asymptotically $N(0,\lambda\tp\Sigma_\psi\lambda)$ almost surely; if $\lambda\tp\Sigma_\psi\lambda=0$ it converges to zero. Applying this to a countable dense set of $\lambda$, and using tightness from the bounded conditional second moments, the conditional law of $L_I$ converges weakly to $N(0,\Sigma_\psi)$ almost surely. For bounded Lipschitz $f$,
\[
\bigl|\E_I f(Z_I)-\E f\{\mathcal N(Z)\}\bigr|\le\bigl|\E_If\{\mathcal N(L_I)\}-\E f\{\mathcal N(Z)\}\bigr|+\|f\|_{\rm BL}\{\epsilon\,\mathrm{Lip}(\mathcal N)+2\pi_n(\epsilon)\},
\]
so the conditional law of $Z_I$ converges in probability to that of $\mathcal N(Z)$.

\emph{Both cases.} Conditional on the data, independently drawn subsets give independent bounded indicators, so their Monte Carlo empirical distribution function has conditional variance at most $1/(4B)$ and the conclusion persists for $B\to\infty$. Continuity and strict increase at the target quantile give $q_{n,b}(1-\gamma)\to q(1-\gamma)$ in probability. For every sufficient pair $(w,A)$, all $T_{wvA}=0$, so
\[
\max_{w,A\in\Acal_w^\Fcal}\sqrt n\widehat Q_w(A)\le\sqrt n\,\mathcal N(\widehat{\mathbf T}_n-\mathbf T)\Rightarrow\mathcal N(Z).
\]
Hence the probability that every sufficient set is retained is at least $P\{\sqrt n\mathcal N(\widehat{\mathbf T}_n-\mathbf T)\le q_{n,b}(1-\gamma)\}\to1-\gamma$. For an insufficient set, $\widehat Q_w(A)\to Q_w(A)>0$, whereas $q_{n,b}(1-\gamma)/\sqrt n\to0$; finiteness of $\Fcal$ gives simultaneous exclusion.
\end{proof}

\begin{proof}[Proof of Proposition~\ref{prop:transfer}]
For any indexed block, write $P=P_w$, $Q=P_W$, and $H=H_v(A)$. The exact decomposition
\[
\widehat P\widehat H\widehat Q-PHQ
=\widehat P(\widehat H-H)\widehat Q
 +(\widehat P-P)H\widehat Q+PH(\widehat Q-Q)
\]
and the contraction property of orthogonal projectors give the block bound
$h_n+L_n(e_{Y,n}+e_{W,n})$. The reverse triangle inequality for operator norms and summation over $v$ prove \eqref{eq:transfer}. Let $E_n=\max_{w,A}|\widehat Q_w(A)-Q_w(A)|$. Then $E_n/t_n\to0$ in probability. On $E_n<t_n$ every zero criterion is retained. If positive criteria exist, on $E_n<\delta_n-t_n$ every positive criterion exceeds $t_n$ and is excluded. Both events have probability tending to one by the stated separation conditions. Because $E_n$ is already a uniform bound, no additional union argument is needed.
\end{proof}

\subsection{Functional inference and efficiency}
\begin{proof}[Proof of Theorem~\ref{thm:effects}]
For a fresh observation and training-fold nuisance estimates held fixed,
\begin{align*}
\widehat\psi_v-\psi_v
&=d_m(1-\I_v/\widehat\pi_v)
+\I_v(\widetilde Y-m_v)(\widehat\pi_v^{-1}-\pi_v^{-1}),\\
P(\widehat\psi_v-\psi_v)&=P(d_m d_\pi/\widehat\pi_v),
\end{align*}
where $\I_v=\I(W=v)$. The squared $L^2$ norm of the first term is bounded by
$2P\{(1+\pi_v/\widehat\pi_v^2)\|d_m\|^2\}$; that of the second is
$P\{\sigma_v^2d_\pi^2/(\pi_v\widehat\pi_v^2)\}$. Equations~\eqref{eq:weighted1}--\eqref{eq:weighted2} therefore give $P\|\widehat\psi_v-\psi_v\|^2=o_P(1)$, and \eqref{eq:weighted3} bounds the norm of the bias by $o_P(n^{-1/2})$. Conditional Chebyshev on each held-out fold gives
$(\Pn-P)(\widehat\psi_v-\psi_v)=o_P(n^{-1/2})$ for its fold average. Summing the finitely many fold contributions yields the asserted expansion. Sufficiency gives $P\psi_v=\E\widetilde Y(v)$, and the weighted residual moments ensure $\E\|\varphi\|^2<\infty$. The separable-Hilbert i.i.d. CLT proves the population limit.

For the conditional sample center, the leading score is
$\zeta=\I_a\varepsilon_a/\pi_a-\I_b\varepsilon_b/\pi_b$.
The indicators are mutually exclusive, so $\Cov(\zeta)=\mathcal C_C$. Also $\E(\zeta\mid X_S)=0$, implying orthogonality to $\tau_S(X_S)-\Delta$ and hence
$\mathcal C_P=\mathcal C_C+\Cov(\tau_S)$.
For the sample potential-outcome center, the leading score is
$\chi=\zeta-(\varepsilon_a-\varepsilon_b)$. Joint sufficiency gives
\[
\E\{\zeta\otimes(\varepsilon_a-\varepsilon_b)\}
=\E\Cov(\varepsilon_a-\varepsilon_b\mid X_S)=:\mathcal D.
\]
The operator $\mathcal D$ is self-adjoint and positive semidefinite; expansion of $\E(\chi\otimes\chi)$ yields $\mathcal C_C-\mathcal D$. The Hilbert CLT for these two scores proves the remaining limits and ordering.

For covariance consistency, the foldwise score $L^2$ bounds imply the empirical mean squared difference between fitted and oracle scores is $o_P(1)$ by conditional Markov's inequality. The inequality
$\|u\otimes u-v\otimes v\|_{\rm tr}\le\|u-v\|(\|u\|+\|v\|)$, Cauchy--Schwarz, and the LLN for the integrable trace-class variable $\varphi\otimes\varphi$ give trace-norm consistency. Finally, if $P(\widehat S=S)\to1$, the selected and fixed-set estimators agree on that event. Their difference multiplied by $\sqrt n$ is then $o_P(1)$ without conditioning on the selection event. This argument requires consistency, not merely a fixed-level confidence collection.
\end{proof}

\begin{proof}[Proof of Theorem~\ref{thm:efficiency}]
Scalarize by any $h\in\Hcal$; ordering all quadratic forms is equivalent to operator ordering. Write scalar conditional means and variances for $\langle h,\widetilde Y(v)\rangle$. If $X_{S'\setminus S}\indep W\mid X_S$, both sets use the same propensity. Put
$a_0=m_{a,S'}-m_{a,S}$ and $b_0=m_{b,S'}-m_{b,S}$. Iterated conditional expectation and total variance give
\[
\langle h,(\mathcal C_S-\mathcal C_{S'})h\rangle
=\E\{a_0^2/\pi_a+b_0^2/\pi_b-(a_0-b_0)^2\}.
\]
The associated two-by-two matrix has nonnegative diagonal entries and determinant
$(1-\pi_a-\pi_b)/(\pi_a\pi_b)\ge0$. This proves the first order for arbitrary finite exposure sets.

For exposure-only added variables, the potential-outcome means and variances do not change. The difference in variances is
\[
\sum_{v=a,b}\E\left[\sigma_{v,S}^2\left\{\E(1/\pi_{v,S'}\mid X_S)-1/\pi_{v,S}\right\}\right]\ge0,
\]
by Jensen and $\pi_{v,S}=\E(\pi_{v,S'}\mid X_S)$. This proves the second order. For a finite recovered family with a unique trace minimizer, the positive gap to every other trace and uniform consistency imply that the empirical minimizer equals it with probability tending to one. The final assertion of Theorem~\ref{thm:effects} then applies.
\end{proof}

\subsection{Dependence}\label{app:dependence}
\begin{proof}[Proof of Lemma~\ref{lem:mixing}]
Let $r'=\operatorname{rank}\Sigma_\Gamma$, choose $A\in\R^{r\times r'}$ of full column rank with $AA\tp=\Sigma_\Gamma$, and put $Z_i=A^+\Gamma_i$. Since $\Gamma_i$ lies in the range of $\Sigma_\Gamma$ almost surely, $\Gamma_i=AZ_i$ and $\sigma(\Gamma_i)=\sigma(Z_i)$. Moreover $\Cov(Z_i,Z_k)=c(d_{ik})I_{r'}$.

First, $\sigma(O_I)\subseteq\sigma(Z_I)\vee\sigma(\varepsilon_I)$, and likewise for $J$. The pair $(Z_I,Z_J)$ is independent of $(\varepsilon_I,\varepsilon_J)$, and $\varepsilon_I$ is independent of $\varepsilon_J$. By the theorem of \citet{csaki1963} on maximal correlation of independent pairs,
\[
\rho\{\sigma(O_I),\sigma(O_J)\}\le\max\bigl[\rho\{\sigma(Z_I),\sigma(Z_J)\},\rho\{\sigma(\varepsilon_I),\sigma(\varepsilon_J)\}\bigr]=\rho\{\sigma(Z_I),\sigma(Z_J)\}.
\]
Second, for jointly Gaussian vectors the maximal correlation equals the largest canonical correlation \citep{kolmogorov1960strong}. Here $\Var(Z_I)=C_{II}\otimes I_{r'}$ and $\Cov(Z_I,Z_J)=C_{IJ}\otimes I_{r'}$, so
\[
\rho\{\sigma(Z_I),\sigma(Z_J)\}=\|C_{II}^{-1/2}C_{IJ}C_{JJ}^{-1/2}\|_{\rm op}\le\frac{\|C_{IJ}\|_{\rm op}}{\lambda_{\min}(C_{II})^{1/2}\lambda_{\min}(C_{JJ})^{1/2}}\le\frac{\|C_{IJ}\|_{\rm op}}{\lambda_*},
\]
since eigenvalues of principal submatrices interlace those of $C_n$. Third, by the Schur test,
$\|C_{IJ}\|_{\rm op}\le\{\max_{i\in I}\sum_{k\in J}c_{ik}\max_{k\in J}\sum_{i\in I}c_{ik}\}^{1/2}\le\sup_i\sum_{k:d_{ik}\ge r}c(d_{ik})$.

Disks of radius $1/2$ centred at the sites are disjoint. Hence for $t\ge1$ the number of sites with $t\le d_{ik}<t+1$ is at most the area of the annulus with radii $t-\tfrac12$ and $t+\tfrac32$ divided by $\pi/4$, namely $16t+8$. Therefore
\[
\sum_{k:d_{ik}\ge r}c(d_{ik})\le C_0\sum_{t\ge\lfloor r\rfloor}(16t+8)e^{-t/\ell}
\le C_0\,A_\ell\sum_{t\ge\lfloor r\rfloor}e^{-t/(2\ell)}\le\frac{C_0A_\ell e^{1/(2\ell)}}{1-e^{-1/(2\ell)}}\,e^{-r/(2\ell)},
\]
with $A_\ell=\sup_{t\ge0}(16t+8)e^{-t/(2\ell)}$. Finally $\alpha\le\rho/4$ for any pair of $\sigma$-fields \citep{bradley2005basic}.
\end{proof}

\begin{proof}[Proof of Proposition~\ref{prop:mehler}]
With $Z_i$ as above, let $f(z)=\E\phi\{F(Az,\varepsilon_1)\}$, so $f(Z_i)=\E\{\phi(O_i)\mid\Gamma\}$ because $\varepsilon_i$ is independent of $\Gamma$ and of $\varepsilon_k$, $k\ne i$. For $i\ne k$, $\E\{\phi(O_i)\phi(O_k)\tp\mid\Gamma\}=f(Z_i)f(Z_k)\tp$ by the same independence. Let $h_\alpha(z)=\prod_jH_{\alpha_j}(z_j)/(\alpha_j!)^{1/2}$ be the normalized Hermite polynomials, an orthonormal basis of $L^2\{N(0,I_{r'})\}$, and let $f_\alpha=\E f(\zeta)h_\alpha(\zeta)$, $\zeta\sim N(0,I_{r'})$. Since $\E\phi=0$, $f_0=0$. The coordinates of $(Z_i,Z_k)$ form $r'$ independent standard bivariate normal pairs with correlation $c=c(d_{ik})$. Mehler's formula gives $\E\{H_a(Z_{ij})H_b(Z_{kj})\}=\delta_{ab}\,a!\,c^a$ \citep[Chapter~1]{nualart2006malliavin}, hence $\E\{h_\alpha(Z_i)h_\beta(Z_k)\}=\delta_{\alpha\beta}c^{|\alpha|}$. The bilinear form $(u,v)\mapsto\E u(Z_i)v(Z_k)$ is bounded on $L^2\times L^2$ by Cauchy--Schwarz, so it may be evaluated termwise on the Hermite expansions of the coordinates of $f$:
\[
\E\{f(Z_i)f(Z_k)\tp\}=\sum_{\alpha\ne0}c^{|\alpha|}f_\alpha f_\alpha\tp=\sum_{m\ge1}c^mM_m,\qquad M_m=\sum_{|\alpha|=m}f_\alpha f_\alpha\tp\succeq0.
\]
The series converges absolutely because $\sum_\alpha\|f_\alpha\|^2=\E\|f(\zeta)\|^2<\infty$ and $|c|\le1$. Further, $\sum_mM_m=\Cov f(\zeta)=\Cov\E(\phi\mid\Gamma_1)\preceq C_P$ by the law of total covariance. For every $\lambda$, $c\mapsto\lambda\tp g(c)\lambda=\sum_mc^m\lambda\tp M_m\lambda$ is a nonnegative combination of convex nondecreasing functions on $[0,1]$, vanishes at zero and is at most $c\,\lambda\tp C_P\lambda$ since $c^m\le c$. Summing \eqref{eq:mehler} over ordered pairs, with Tonelli's theorem for the nonnegative terms of each quadratic form, gives \eqref{eq:Csp}.

For \eqref{eq:undercover}, $\widehat C_P\to C_P$ in probability by Lemma~\ref{lem:pairlln} below with $h=0$, using the fourth moment. The interval $\langle h,\widehat\Delta\rangle\pm z_{1-a/2}(h\tp\widehat C_Ph/n)^{1/2}$ then covers with probability tending to $P(|N(0,1)|\le z_{1-a/2}r_h)$.
\end{proof}

\begin{lemma}[Pair averages]\label{lem:pairlln}
Under Assumption~\ref{ass:latent}, let $\phi$ satisfy $\E\|\phi(O_1)\|^4<\infty$, let $h\ge0$, and let $B_n\subseteq\{(i,k):d_{ik}\le h\}$ with weights $|w_{ik}|\le1$. Then
\[
\frac1n\sum_{(i,k)\in B_n}w_{ik}\bigl[\phi(O_i)\phi(O_k)\tp-\E\{\phi(O_i)\phi(O_k)\tp\}\bigr]\to0\quad\text{in }L^2 .
\]
\end{lemma}
\begin{proof}
Let $\xi_i=\sum_{k:(i,k)\in B_n}w_{ik}\phi(O_i)\phi(O_k)\tp$. By the packing bound, the number of $k$ with $d_{ik}\le h$ is at most $N_h=4(h+\tfrac32)^2$, and Cauchy--Schwarz gives $\|\lambda\tp\xi_i\lambda\|_{L^2}\le N_h\|\lambda\|^2\{\E\|\phi\|^4\}^{1/2}=:\kappa$. The variable $\xi_i$ is measurable with respect to the observations within distance $h$ of $s_i$. For $d_{ij}>2h$ these sets are separated by at least $d_{ij}-2h$, so by the covariance inequality defining $\rho$ and Lemma~\ref{lem:mixing},
$|\Cov(\lambda\tp\xi_i\lambda,\lambda\tp\xi_j\lambda)|\le\kappa^2\min\{1,K_0e^{-(d_{ij}-2h)/(2\ell)}\}$.
By the packing bound, $\sum_j\min\{1,K_0e^{-(d_{ij}-2h)/(2\ell)}\}$ is bounded uniformly in $i$ and $n$. Hence the variance of $n^{-1}\sum_i\lambda\tp\xi_i\lambda$ is $O(1/n)$. Polarization extends this to all entries.
\end{proof}

\begin{proof}[Proof of Theorem~\ref{thm:depclt}]
For fixed $\vartheta$, the variables $g(O_i,\vartheta)$ are identically distributed. By the covariance inequality defining $\rho$, Lemma~\ref{lem:mixing} and the packing bound,
\[
\Var\Bigl\{\frac1n\sum_i\lambda\tp g(O_i,\vartheta)\Bigr\}\le\frac Cn\E\{\lambda\tp g(O_1,\vartheta)\}^2,
\]
so $n^{-1}\sum_ig(O_i,\vartheta)-\E g(O_1,\vartheta)\to0$ in $L^2$. The same holds for $L_i=\sup_\Theta\|\partial_\vartheta g(O_i,\vartheta)\|$. Since $\|n^{-1}\sum_i\{g(O_i,\vartheta)-g(O_i,\vartheta')\}\|\le(n^{-1}\sum_iL_i)\|\vartheta-\vartheta'\|$, a finite-net argument on the compact $\Theta$ gives a uniform law of large numbers for $g$. The same argument applies to $\partial_\vartheta g$, using its continuity and the dominated-convergence modulus of continuity. Uniqueness of the zero and compactness give $\widehat\vartheta\to\vartheta_0$ \citep[Theorem~5.9]{vandervaart1998asymptotic}. The covariance bound also gives $\bar g_n(\vartheta_0)=n^{-1}\sum_ig(O_i,\vartheta_0)=O_P(n^{-1/2})$. A mean-value expansion of $0=\bar g_n(\widehat\vartheta)$ around $\vartheta_0$, with rows evaluated at intermediate points and the uniform law for $\partial_\vartheta g$, gives
\[
\widehat\vartheta-\vartheta_0=-G^{-1}\bar g_n(\vartheta_0)+o_P(n^{-1/2}).
\]
By Assumption~\ref{ass:depaipw} the target component of $\vartheta_0$ equals $\theta$. For the three examples this follows from Theorems~\ref{thm:causalid} and~\ref{thm:mediation} and the incremental calculation of Section~\ref{sec:stochastic}, applied to the marginal law of one unit, which is common to all units. Hence the asserted linearization holds with $\phi$.

For the central limit theorem, fix $\lambda$ and set $Z_{i,n}=\lambda\tp\phi(O_i)$. These are measurable functions of $O_i$. Lemma~\ref{lem:mixing} bounds the strong-mixing coefficient between the $\sigma$-fields of any two index sets at distance at least $r$ by $K_0e^{-r/(2\ell)}/4$, uniformly in their cardinalities, so the mixing coefficients of \citet{jenish2009central} satisfy $\alpha_{k,l}(r)\le K_0e^{-r/(2\ell)}/4$ for all $k,l$, including $l=\infty$. The moment bound gives uniform $L^{2+\kappa}$ integrability, and $\liminf_nn^{-1}\Var\sum_iZ_{i,n}\ge\liminf\lambda_{\min}(\Csp_n)\|\lambda\|^2>0$. All conditions of the central limit theorem of \citet{jenish2009central} for arrays of $\alpha$-mixing fields on $\R^2$ with minimal distance one are met. Thus $(\lambda\tp\Csp_n\lambda)^{-1/2}n^{-1/2}\sum_iZ_{i,n}\Rightarrow N(0,1)$. The sequence $\Csp_n$ is bounded by Proposition~\ref{prop:mehler} and the packing bound, and bounded away from singularity by assumption. Passing to subsequences along which $\Csp_n$ converges and applying the Cram\'er--Wold device gives the matrix-normalized statement.
\end{proof}

\begin{proof}[Proof of Theorem~\ref{thm:covest}]
\emph{Plug-in errors.} Since $\widehat\phi_i-\phi_i=\phi(O_i;\widehat\vartheta)-\phi(O_i;\vartheta_0)$ is bounded by $L(O_i)\|\widehat\vartheta-\vartheta_0\|$ with $\E L^2<\infty$, $n^{-1}\sum_i\|\widehat\phi_i-\phi_i\|^2=O_P(n^{-1})$. For any $B_n$ as in Lemma~\ref{lem:pairlln}, Cauchy--Schwarz and the bound $N_h$ on neighbours give
\[
\frac1n\sum_{(i,k)\in B_n}\|\widehat\phi_i\widehat\phi_k\tp-\phi_i\phi_k\tp\|\le\frac{2N_h}{n}\Bigl(\sum_i\|\widehat\phi_i-\phi_i\|^2\Bigr)^{1/2}\Bigl(\sum_i\|\widehat\phi_i\|^2+\|\phi_i\|^2\Bigr)^{1/2}=o_P(1).
\]
Centring at $\overline{\widehat\phi}=O_P(n^{-1/2})$ changes these averages by $o_P(1)$, and the corrections of order $1/n$ in the regressors and responses are $o_P(1)$. With probability tending to one, $\widehat\ell\le3\ell/2$. On that event $|\widehat c_{ik}^{\,m}-c_{ik}^m|\le m\,d_{ik}|\widehat\ell^{-1}-\ell^{-1}|e^{-2d_{ik}/(3\ell)}$, whose sums over $k$ are bounded uniformly in $i$ by the packing bound. Hence $\widehat S_m-S_{m,n}\to0$ and $n^{-1}\sum_{i\ne k}|\widehat c_{ik}-c_{ik}|\to0$ in probability.

\emph{(a).} With $m_0$ finite chaos, $\E\phi_i\phi_k\tp=\sum_{m\le m_0}c_{ik}^mM_m$ exactly. By Lemma~\ref{lem:pairlln} with weights $c_{ik}^m$, the normalized cross-products $|\mathcal P_n|^{-1}\sum_{\mathcal P_n}a_{ik}\otimes\widetilde\phi_i\widetilde\phi_k\tp$ differ from $|\mathcal P_n|^{-1}\sum_{\mathcal P_n}a_{ik}a_{ik}\tp M$ by $o_P(1)$, since $|\mathcal P_n|/n$ is bounded away from zero. The estimated Gram matrices differ from the population ones by $o_P(1)$, and the latter have eigenvalues bounded away from zero. Hence the normalized least-squares objective converges, uniformly on bounded sets, to a strictly convex quadratic whose unique minimizer $(M_1,\ldots,M_{m_0})$ lies in the constraint set. Its curvature is bounded below by the smallest eigenvalue of the Gram matrix, so the constrained minimizers converge to $(M_1,\ldots,M_{m_0})$. Since $\widehat C_P-C_P\to0$ and $S_{m,n}$ is bounded, $\widehat V_{\rm M}-\Csp_n\to0$.

\emph{(b).} Let $x_j=|B_j|^{-1}\sum_{B_j}c_{ik}$ and $y_j=|B_j|^{-1}\sum_{B_j}g(c_{ik})$ be the population bin means. By Proposition~\ref{prop:mehler}, $\E\phi_i\phi_k\tp=g(c_{ik})$, so Lemma~\ref{lem:pairlln} and the plug-in bounds give $\bar c_j-x_j\to0$ and convergence of the uncorrected bin means of the centred products to $y_j$ in probability. The correction $\widehat V_{\rm ch}/n$ is $O_P(1/n)$: in the fixed-point equation $\widehat V_{\rm ch}$ enters with total weight $2n^{-2}\sum_{i<k}\omega_{ik}$, where $\omega_{ik}\le1$ is the weight that pair $(i,k)$ places on interior knots. Pairs within distance $h$ number $O(n)$ by the packing bound, and every other pair has $\omega_{ik}\le\widehat c_{ik}/\bar c_{(1)}$, whose sum is $O_P(nS_{1,n})$ by the plug-in bounds. The weight is therefore $O_P(1/n)$, the map is a contraction for large $n$, and its fixed point exists, is unique and is $O_P(1)$. Projection onto the positive semidefinite cone is a contraction in the Frobenius norm that fixes the positive semidefinite matrices $y_j$, so $\|\widehat g_j-y_j\|$ is at most the error before projection. Hence $\widehat g_j-y_j\to0$ in probability. Let ${\rm ch}_n$ be the piecewise-linear interpolant through $(0,0)$, the points $(x_j,y_j)$ and $(1,C_P)$. Fix $\lambda$ and set $s(c)=\lambda\tp g(c)\lambda$, a convex function with $s(0)=0$. By Jensen's inequality $\lambda\tp y_j\lambda\ge s(x_j)$, and $\lambda\tp C_P\lambda\ge s(1)$. On each segment, the linear interpolant of values that dominate $s$ at the end points dominates the chord of $s$, and hence $s$, by convexity. Therefore $\lambda\tp{\rm ch}_n(c)\lambda\ge s(c)$ on $[0,1]$, so $E_n\succeq0$ and
\[
C_P+\frac2n\sum_{i<k}{\rm ch}_n(c_{ik})=\Csp_n+E_n .
\]
Since $d_{ik}\ge1$, all $x_j\le e^{-1/\ell}<1$. The knots are separated by assumption and bounded away from $0$ and $1$. Hence $\sup_{0<c\le1}\|\widehat{\rm ch}(c)-{\rm ch}_n(c)\|/c\to0$ in probability: on $[0,\min(\bar c_{(1)},x_{(1)})]$ both functions are linear through the origin with converging slopes, and above it $c$ is bounded below. The Lipschitz constant of $\widehat{\rm ch}$ is $O_P(1)$. Therefore
\[
\frac2n\sum_{i<k}\|\widehat{\rm ch}(\widehat c_{ik})-{\rm ch}_n(c_{ik})\|\le O_P(1)\,\frac1n\sum_{i\ne k}|\widehat c_{ik}-c_{ik}|+o_P(1)\,S_{1,n}=o_P(1).
\]
Together with $\widehat C_P-C_P\to0$ this gives $\widehat V_{\rm ch}=\Csp_n+E_n+o_P(1)$. Both assertions of (b) follow, since $E_n\succeq0$ and the final positive semidefinite projection can only increase a symmetric matrix in the Loewner order.

\emph{(c).} For a functional, $\lambda\tp\sqrt n(\widehat\Delta-\Delta)/(\lambda\tp\widehat V\lambda)^{1/2}$ is the product of a statistic converging to $N(0,1)$ (Theorem~\ref{thm:depclt}) and the factor $(\lambda\tp\Csp_n\lambda/\lambda\tp\widehat V\lambda)^{1/2}$. This factor converges to one under (a) and is asymptotically at most one under (b). For the $L^2$ statistic $n\|\widehat\Delta-\Delta\|_w^2$, the null law computed from $\widehat V$ is that of $\|Z_1+Z_2\|_w^2$ with independent centred Gaussian $Z_1\sim N(0,\Csp_n)$ and $Z_2\sim N(0,\widehat V-\Csp_n)$, up to $o_P(1)$. By Anderson's inequality for symmetric convex sets \citep{anderson1955integral}, its quantiles are at least those of $\|Z_1\|_w^2$.

\emph{(d).} Let $\kappa=\ell'/\ell\ge1$ and $c'_{ik}=e^{-d_{ik}/\ell'}$, so that $c_{ik}=(c'_{ik})^\kappa$. Define $h(x)=g(x^\kappa)$ on $[0,1]$. For every $\lambda$, $x\mapsto\lambda\tp h(x)\lambda$ is the composition of the convex nondecreasing function $c\mapsto\lambda\tp g(c)\lambda$ with the convex nondecreasing map $x\mapsto x^\kappa$, hence convex and nondecreasing; moreover $h(0)=0$ and $h(1)=g(1)\preceq C_P$. Since $\E\phi_i\phi_k\tp=h(c'_{ik})$, the proof of (b) applies verbatim with $(h,c')$ in place of $(g,c)$, and $\widehat\ell\to\ell'$ replaces $\widehat\ell\to\ell$. It gives $\widehat V_{\rm ch}=\Csp_n+E'_n+o_P(1)$ with $E'_n\succeq0$ the chord error of $h$, because $C_P+n^{-1}\sum_{i\ne k}h(c'_{ik})=\Csp_n$. For $\kappa<1$ the map $x^\kappa$ is concave and the argument fails, which is why only overstatement of the range is covered. The Wald and $L^2$ conclusions then follow as in (c).
\end{proof}

\begin{proof}[Proof of Theorem~\ref{thm:blockhalf}]
Let $G=G_n$ be even, $U$ the union of a uniformly random half of the blocks, $\psi_i=\psi_T(O_i)$ and $\epsilon_g=2\I(g\subseteq U)-1$. Since $|U|=n/2$,
\[
\sqrt n(\widehat{\mathbf T}_U-\widehat{\mathbf T}_n)=L_U+\sqrt n(R_U-R_n),\qquad L_U=n^{-1/2}\sum_{g=1}^G\epsilon_gV_g .
\]
As in the proof of Theorem~\ref{thm:subsample}, the remainder assumption and Markov's inequality make the conditional probability that $\sqrt n\|R_U-R_n\|>\epsilon$ tend to zero in probability.

Conditionally on the data, $\E(\epsilon_g)=0$, $\E\epsilon_g^2=1$ and $\E\epsilon_g\epsilon_h=-1/(G-1)$ for $g\ne h$. Hence $L_U$ has conditional mean zero and conditional covariance
\[
\frac{G}{G-1}\,\frac1n\sum_g(V_g-\bar V)(V_g-\bar V)\tp
=\frac{G}{G-1}\Bigl\{\frac1n\sum_gV_gV_g\tp-\frac{1}{nG}\Bigl(\sum_gV_g\Bigr)\Bigl(\sum_gV_g\Bigr)\tp\Bigr\}.
\]
The last term is $O_P(1/G)$ because $\sum_gV_g=O_P(\sqrt n)$. We show $n^{-1}\sum_gV_gV_g\tp-\Csp_{T,n}\to0$ in probability. By Proposition~\ref{prop:mehler} applied to $\psi_T$, $\|\Cov(\psi_i,\psi_k)\|\le c_{ik}\|C_{P,T}\|$ for $i\ne k$. Hence
\[
\Bigl\|\E\frac1n\sum_gV_gV_g\tp-\Csp_{T,n}\Bigr\|\le\frac{\|C_{P,T}\|}{n}\sum_i\sum_{k\notin g(i)}c_{ik},
\]
where $g(i)$ is the block containing $i$. For a unit at distance at least $t$ from the boundary of its block, the inner sum is at most $K_0\lambda_*e^{-t/(2\ell)}$ by the proof of Lemma~\ref{lem:mixing}. By the boundary condition of Assumption~\ref{ass:blocks}, at most a fraction $C_B(1+t)m_n^{-1/2}$ of units lie within distance $t$ of a unit in another block, and for them the inner sum is bounded. Taking $t=m_n^{1/4}$ shows that the bias vanishes. For the variance, fix $\lambda$ and let $X_g=(\lambda\tp V_g)^2$. By the covariance inequality for $\rho$, the block moment bound and Lemma~\ref{lem:mixing}, $|\Cov(X_g,X_h)|\le Cm_n^2\min\{1,K_0e^{-r_{gh}/(2\ell)}\}$, where $r_{gh}$ is the distance between blocks. Each block contains $m_n$ sites with separation one, hence covers area at least $cm_n$, and lies in a square of side $C_Bm_n^{1/2}$; so at most $C'(1+t/m_n^{1/2})^2$ blocks lie within distance $t$ of a given block, and $\sum_h\min\{1,K_0e^{-r_{gh}/(2\ell)}\}$ is bounded uniformly. Thus $\Var(n^{-1}\sum_gX_g)\le C''Gm_n^2/n^2=C''m_n/n\to0$.

For H\'ajek's condition on the block population, $\E\max_g\|V_g\|^4\le\sum_g\E\|V_g\|^4\le CGm_n^2=Cnm_n$, so $\max_g\|V_g\|^2/n=O_P\{(m_n/n)^{1/2}\}\to0$. The sampled number of blocks and its complement both equal $G/2\to\infty$. As in the proof of Theorem~\ref{thm:subsample}, the conditional law of $L_U$ therefore converges in probability to $N(0,\Csp_T)$, and that of $Z_U$ to the law of $\mathcal N(Z)$. Finally, $\sqrt n(\widehat{\mathbf T}_n-\mathbf T)\Rightarrow N(0,\Csp_T)$: the linearization is assumed, and the central limit theorem follows as in the proof of Theorem~\ref{thm:depclt}. The quantile and inclusion arguments of Theorem~\ref{thm:subsample} complete the proof.
\end{proof}

\section*{Declaration of generative AI and AI-assisted technologies in the manuscript preparation process}
During the preparation of this work the author used Claude (Anthropic) and ChatGPT/Codex (OpenAI) to assist with drafting and editing text, deriving and checking proofs, writing analysis code and preparing figures. The figures are data visualizations produced directly from the stored numerical results by reproducible analysis code. After using these tools, the author reviewed and edited the content as needed and takes full responsibility for the content of the published article.

\bibliographystyle{elsarticle-num-names}
\bibliography{references}
\end{document}